%% file: main.tex
\documentclass[conference,compsoc]{IEEEtran}
\ifCLASSOPTIONcompsoc
  \usepackage[nocompress]{cite}
\else
  \usepackage{cite}
\fi
\usepackage[backref=page]{hyperref}

\usepackage{subfloat}
\usepackage{adjustbox}
\usepackage{multirow}
\ifCLASSINFOpdf
  \graphicspath{{./images/}{./figures/}}
  \DeclareGraphicsExtensions{.pdf,.jpeg,.jpg,.png}
\else
  \graphicspath{{./images/}{./figures/}}
  \DeclareGraphicsExtensions{.eps}
\fi
\usepackage{mdframed}

\usepackage{url}

\usepackage{booktabs}
\newcommand{\ra}[1]{\renewcommand{\arraystretch}{#1}}

\usepackage{multirow}
\usepackage{proofread}

\input{setup_comments.tex} 

\newcommand\ourdomainshort{{\sc MCS}}

\newcommand\DistribSys{distributed systems}

\newcommand{\AtLarge}{{\sc AtLarge}}

\newcommand{\theDesigner}{{the designer}}

\newcommand{\vcutS}{\vspace*{-0.15cm}}
\newcommand{\vcutM}{\vspace*{-0.25cm}}
\newcommand{\vcutL}{\vspace*{-0.5cm}}

\newcommand\vspsmall{\vspace*{-0.15cm}}
\newcommand\vspmed{\vspace*{-0.25cm}}

\usepackage{tcolorbox}
\tcbuselibrary{skins}

\definecolor{BgBlue}{HTML}{336699}
\definecolor{BgGreen}{HTML}{339966}
\definecolor{BgYellow}{HTML}{664422}
\definecolor{BgOrange}{HTML}{994422}

\newcommand\vision[1]{
	\begin{tcolorbox}[colback=BgYellow!15, colframe=BgYellow!90!black,enhanced,drop fuzzy shadow,sharp corners]
		\textbf{Vision}: #1
	\end{tcolorbox}
}

\newcommand\definition[1]{
	\begin{tcolorbox}[colback=orange!15, colframe=orange!90!black,enhanced,drop fuzzy shadow,sharp corners]
		\textbf{Definition}: #1
	\end{tcolorbox}
}

\newcounter{principleid}
\newcommand{\rprinciple}[1]{\refstepcounter{principleid}\label{#1}}

\newcommand\principle[2]{
	\rprinciple{#1}
	\begin{tcolorbox}[colback=BgBlue!15, colframe=BgBlue!90!black,enhanced,drop fuzzy shadow,sharp corners, boxsep=2pt,left=4.5pt,right=4.5pt,top=2pt,bottom=2pt]
		\textbf{P\arabic{principleid}}: #2
	\end{tcolorbox}
}

\newcommand\refpr[2]{\textbf{P\ref{#1}}}

\newcounter{challengeid}
\newcommand{\rchallenge}[1]{\refstepcounter{challengeid}\label{#1}}

\newcommand\challenge[3]{ 
	\rchallenge{#1}
	\begin{tcolorbox}[colback=BgGreen!15, colframe=BgGreen!90!black,enhanced,drop fuzzy shadow,sharp corners,boxsep=2pt,left=4.5pt,right=4.5pt,top=2pt,bottom=2pt]
		\textbf{C\arabic{challengeid}}: #3
	\end{tcolorbox}
}

\usepackage{mathabx}

\begin{document}
%
\title{The AtLarge Vision on the Design of Distributed Systems and Ecosystems: \\Extended Technical Report}

\author{\IEEEauthorblockN{Alexandru Iosup}
\IEEEauthorblockA{Department of Computer Science,\\
Faculty of Sciences, VU Amsterdam,\\
The Netherlands\\
A.Iosup@vu.nl}
\and
\IEEEauthorblockN{Laurens Versluis}
\IEEEauthorblockA{Department of Computer Science,\\
Faculty of Sciences, VU Amsterdam,\\
The Netherlands\\
L.F.D.Versluis@vu.nl}
\and
\IEEEauthorblockN{The AtLarge Team\IEEEauthorrefmark{1}}
\IEEEauthorblockA{VU Amsterdam and\\
Delft University of Technology,\\
The Netherlands,\\
\url{http://atlarge.science}}
\IEEEcompsocitemizethanks{
\IEEEcompsocthanksitem The other AtLarge team members co-authoring this article are: 
Animesh Trivedi, 
Erwin van Eyk, Lucian Toader, Vincent van Beek, Giulia Frascaria, Ahmed Musaafir, and Sacheendra Talluri.}
}


%


\maketitle

\input{content/00-abstract.tex}



\input{content/01-intro.tex}


\input{content/02-motivation.tex}

\input{content/03-atlarge-framework.tex}

\input{content/04-design-principles.tex}

\input{content/05-design-challenges.tex}

\input{content/06-design-exp.tex}

\input{content/01-relatedwork.tex}

\input{content/01-conclusion.tex}

\input{content/00-ack.tex}

\bibliographystyle{IEEEtran}
\bibliography{main_beautified,main_beautified_beautified}

\end{document}

%% file: setup_comments.tex
\usepackage{xcolor}  
\definecolor{OwnAzure}{HTML}{336699}
\definecolor{OwnCerulean}{HTML}{CAE2FE}
\definecolor{OwnOliveGreen}{HTML}{556B2F}
\definecolor{KamPurple}{HTML}{907C97}
\usepackage[colorinlistoftodos,prependcaption,textsize=tiny]{todonotes}
\usepackage{xargs}                      
\newcommandx{\todolarge}[2][1=]{\todo[inline,size=\large,linecolor=OwnAzure,backgroundcolor=OwnCerulean,bordercolor=OwnAzure,#1]{#2}}
\newcommandx{\todoai}[2][1=]{\todo[inline,linecolor=OwnAzure,backgroundcolor=OwnCerulean,bordercolor=OwnAzure,#1]{#2}}

\newcommandx{\addref}[2][1=]
{\todo[inline,linecolor=blue,backgroundcolor=blue!50,bordercolor=blue,#1]{Add reference. #2}}
\newcommandx{\unsure}[2][1=]{\todo[inline, linecolor=red,backgroundcolor=red!25,bordercolor=red,#1]{#2}}
\newcommandx{\change}[2][1=]{\todo[inline, linecolor=blue,backgroundcolor=blue!25,bordercolor=blue,#1]{#2}}
\newcommandx{\info}[2][1=]{\todo[linecolor=OwnOliveGreen,backgroundcolor=OwnOliveGreen!25,bordercolor=OwnOliveGreen,#1]{#2}}
\newcommandx{\improvement}[2][1=]{\todo[linecolor=Plum,backgroundcolor=Plum!25,bordercolor=Plum,#1]{#2}}
\newcommandx{\thiswillnotshow}[2][1=]{\todo[disable,#1]{#2}}

\definecolor{darkred}{rgb}{0.5,0,0}
\definecolor{darkgreen}{rgb}{0,0.5,0}
\definecolor{darkblue}{rgb}{0,0,0.5}



%% file: content/00-abstract.tex
\begin{abstract}
High-quality designs of distributed systems and services are essential for our digital economy and society. 
Threatening to slow down the stream of working designs, 
we identify the mounting pressure of scale and complexity of \mbox{(eco-)system}, of ill-defined and wicked problems, and of unclear processes, methods, and tools.
We envision design itself as a core research topic in distributed systems, to understand and improve the science and practice of distributed (eco-)system design. 
Toward this vision, 
we propose the \AtLarge{} design framework, accompanied by a set of 8 core design principles. We also propose 10 key challenges, which we hope the community can address in the following 5 years.
In our experience so far, the proposed framework and principles are practical, and lead to pragmatic and innovative designs for large-scale 
distributed systems.
\end{abstract}

%% file: content/01-intro.tex
\section{Introduction}
\label{sec:introduction}\label{sec:intro}

\begin{figure}[!t]
 \centering
 \includegraphics[width=1.0\columnwidth]{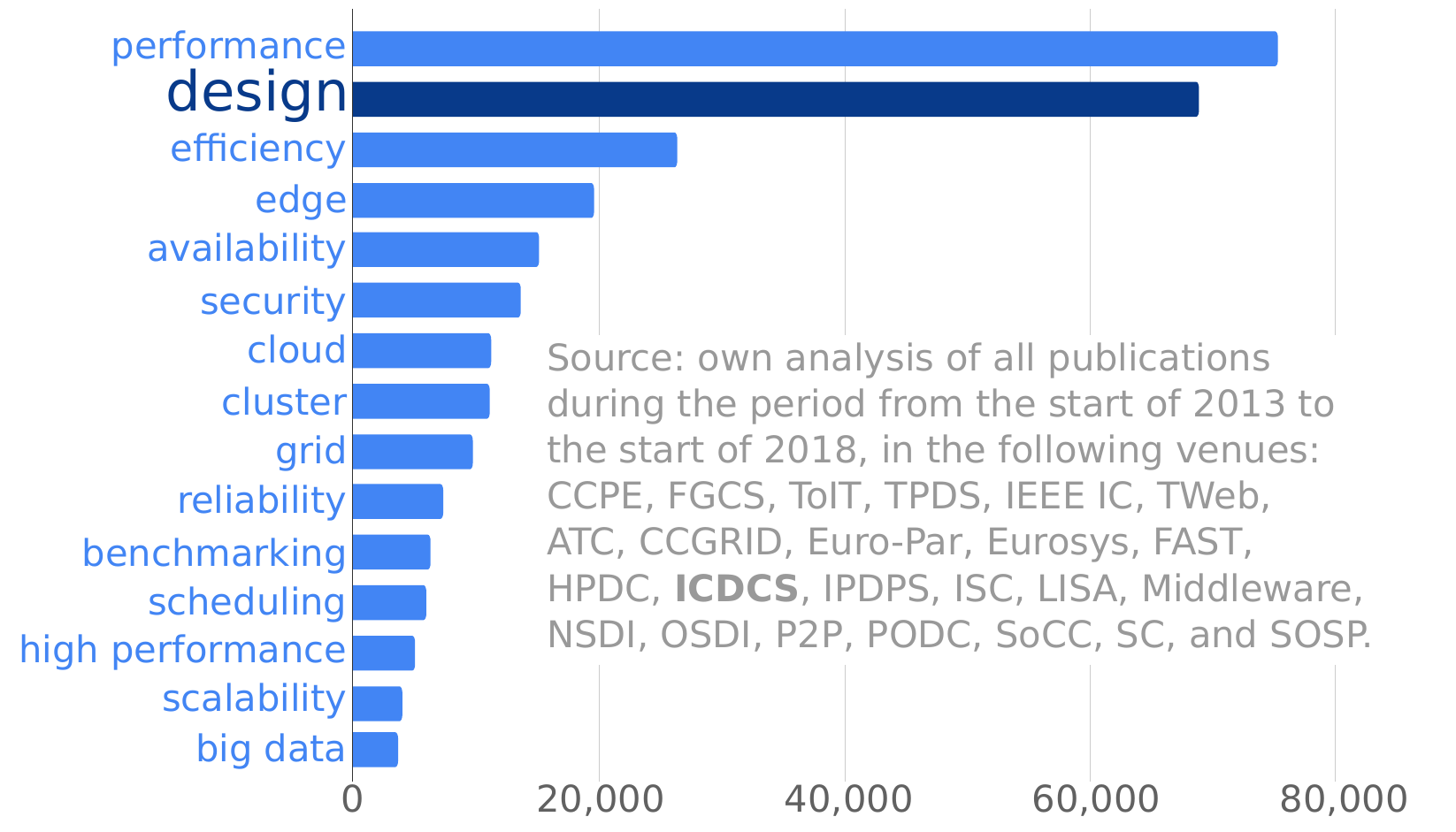} 
 \caption{Presence of selected keywords in top systems venues.}
 \label{fig:designkeyword}
 \vspmed
\end{figure}

A continuous stream of designs, of computer-based services and of the distributed systems on which they run,is expected in our knowledge-based society. We use daily many distributed ecosystems~\cite{DBLP:conf/icdcs/IosupUVAEHTBT18} whose designs appeared only a relatively short while ago, e.g., of GAFAM and BAT~\cite{issues:Economist18a}, and expect new designs that will lead to considerable economic growth and productivity~\cite{tr:ec14cloud,tr:eu16bigdata,tr:gartner17cloud}. As Figure~\ref{fig:designkeyword} indicates, design is a common keyword in top scientific and industry venues, including ICDCS. Yet, as we show in this work, we should not take design for granted, and we should not consider that the current approaches will continue to deliver good results. Design problems keep getting more difficult to formulate, and their solutions more difficult to find and reason about. Existing design processes, from merely relying on intuition to  classic~\cite{design:sweng:Royce70,DBLP:journals/computer/Boehm88,book:design:BlaauwB97} to emerging~\cite{DBLP:journals/csur/RamsinP08,book:design:Burns18}, have significant shortcomings for designing distributed ecosystems~\cite{design:book/Brooks10,DBLP:conf/icdcs/IosupUVAEHTBT18}.
Instead, to address this grand challenge of the distributed systems community, we propose a vision toward establishing new theoretical and practical means to produce pragmatic and innovative designs.

\definition{
``Design is the intentional solution of a problem, by the creation of plans for a new sort of thing, where the plans would not be immediately seen, by a reasonable person, as an inadequate solution.''~\cite[Loc.345]{book:design:Parsons15}. 
{\it Pragmatic design} can be implemented, and evidence shows it can run in production-like settings.
{\it Innovative design} ``consists in novel solutions''~\cite[Loc.2353]{design:book/Arthur09}.
}

We are interested in a particular kind of design, 
for {\it massivizing computer systems~(MCS)}~\cite{DBLP:conf/icdcs/IosupUVAEHTBT18}, that is, for production-ready distributed systems and ecosystems. As in our previous work, we see distributed ecosystems as composites of interconnected (distributed) systems and, recursively, ecosystems. Ecosystems fulfill functional requirements ({\it FR}s), such as responding to service-queries, or batch-processing big data and computation, and non-functional requirements ({\it NFR}s), such as predictable high performance and availability. They do so subject to Service Level Agreements~({\it SLA}s), and in doing so they experience {\it dynamics}, such as provisioning and releasing resources from an external cloud, and give rise to various {\it phenomena} that are difficult to foresee at design time, such as performance variability.

\vision{
We envision a world of
distributed ecosystems, based on pragmatic and innovative \ourdomainshort{} designs, created by diverse designers using design philosophy, processes, patterns, and tools, 
together
with scientists, engineers, and the society itself. 
}

We see design as a major challenge for the field of \ourdomainshort{}, and raise about it two key questions.
{\it How to find good designs and even good problems?} 
The ever-increasing complexity of the
field---contrast the relatively simple design of the 
earlier distributed system BitTorrent with the current ecosystems at Google, which can require the orchestration of hundreds of services and systems to produce meaningful results~\cite{book:google:SRE16,book:google:SRW18}---
makes it unlikely that good designs can be achieved from mere sparks of intuition of lonely designers, without good process and collaboration. 
Not only solving, but also finding problems is increasingly more difficult, and, for ecosystems, finding who should solve them; 
in contrast, in the 1960s, the core systems problems were well-known, and a small architectural team could direct the large team working on the IBM system 360 family~\cite{book:design:BlaauwB97}.

{\it How to design the processes and create the bodies of knowledge that increase the likelihood of good \ourdomainshort{} design?}
It is challenging to select the design elements elements 
that could lead to a high likelihood of good \ourdomainshort{} designs~\cite{conf/hpdc/DesignProcess19}, 
from the hundreds of 
design patterns~\cite{book:design:Abbott15,book:design:Erl15,book:design:Burns18} and
practical steps~\cite{book:design:LidwellHB10,book:design:Delft14,book:design:MartinH12}, and from the many development processes such as rational and agile~\cite{DBLP:journals/csur/RamsinP08}.
Students and even practitioners have rarely studied these systematically, which compounds the problem.
But even if the designer would have the experience and knowledge to select, these design elements make many unreasonable assumptions about how designers actually work~\cite[Ch.3]{design:book/Brooks10}, disregard modern design theory~\cite{book:design:Dorst17}~\cite[Ch.1-2]{book:design:Parsons15}, and focus not on \ourdomainshort{} but on engineering software services~\cite{book:design:Burns18}, software~\cite{book:design:Abbott15,book:design:Bass15}, and hardware~\cite{book:design:BlaauwB97,book:compsys:HennessyP17}.

Our vision aims to place design as a core research topic in distributed systems and ecosystems. 
We do not merely aim to provide a set of {\it design patterns}, which is a staple of software~\cite{book:design:GOF94} and of service~\cite{book:design:Burns18} design but not necessarily the key to design success in distributed systems or even in architecture~\cite[Loc.572]{design:book/Lawson05}\footnote{The approach based on design patterns in architecture~\cite{book:design:Alexander77}, which has inspired generations of software engineers~\cite{book:design:GOF94}, was quickly dismissed by the architecture community, including by its author, as too limiting.} 
We also want to steer away from heavyweight design processes, which stifle good design~\cite[p.233]{design:book/Brooks10}~\cite{design:book/Lawson05}.
We aim to provide a framework for design, 
from understanding how to think about design in this field
to finding and solving \ourdomainshort{} design-problems, 
from design of distributed ecosystems to design supporting experiments of and publications about them,
with a five-fold contribution:

\begin{enumerate}
	\item We are the first to explicitly posit that design is a key area of research in \DistribSys{}, and especially in \ourdomainshort{}~(in Section~\ref{sec:motivation}).
	As support, we offer qualitative and quantitative evidence. 
    
    \item We propose the \AtLarge{} framework for design (Section~\ref{sec:core}). 
    The framework 
    starts from the central premise that design has a fundamentally different nature from science and engineering, which has not been formulated for the distributed systems field.
    It includes novel elements, focusing on \ourdomainshort{}, about design thinking, problem-finding, and reporting, and for problem-solving it leverages
    the basic design cycle we have previously developed~\cite{conf/hpdc/DesignProcess19}.

    \item We propose 8 core principles of \ourdomainshort{} design (Section~\ref{sec:core:principles}). 
    The core principles address four main categories, around the central premise, and systems, peopleware~\cite{concepts:book/DeMarcoL12}, and methodological aspects.

    \item We identify 10 current 
    challenges raised by \ourdomainshort{} design~(in Section~\ref{sec:challenges}). 
    The challenges are grouped into the same four main categories as the core principles---central premise, systems, peopleware~\cite{concepts:book/DeMarcoL12}, and methodological---and give a broad scope of what the field could address in the next 5 years. 
    Although, in doing so, the community and our own work will supersede the framework elements presented here, we envision the general structure of the framework will be long-lasting. 
    
    \item We show evidence, trough real-world experiments, of how the \AtLarge{} design framework can be pragmatic yet lead to innovative designs~(in Section~\ref{sec:use}), and compare the framework with a multi-disciplinary body of related work~(Section~\ref{sec:related}). 
    
\end{enumerate}

%% file: content/02-motivation.tex
\section{Why Focus on \ourdomainshort{} Design?} \label{sec:whydesign} \label{sec:motivation}

We argue in this section for the timely and important need to focus on \ourdomainshort{} design. Not only is (good) design needed~(Section~\ref{sec:motivation:withoutdesign}), 
but we identify an {\it increasing} need for good design~(Section~\ref{sec:needfordesign}) and designers (Section~\ref{sec:motivation:needfordesigners}).
We also analyze what good design needs to address, that is, complex challenges from system design (Section~\ref{sec:motivation:traditional}) and from \ourdomainshort{} design (Section~\ref{sec:motivation:mcs}).

\vcutS{}
\subsection{Without (Good) Design} \label{sec:motivation:withoutdesign}
\vcutM{}

Similarly to how Brooks dismissed the idea that organizations can cope with increasing technical debt just by adding more person-months, in this work we want to dismiss the idea that organizations can cope with increasing system complexity (to parallel Brooks, {\it design debt}) just by hoping good design will simply emerge.

The consequences of not having good designs are well-known, but difficult to quantify. 
Lackluster design costs money, causes systems to under-perform and sometimes to fail, and delays the arrival of needed systems in the market. Organizations prioritizing working systems over good design effectively defer the moment when they will have to actually solve the problem. In many cases, careful monitoring and capable engineering teams (e.g., sysadmins or site reliability engineers) can help resolve the problems, and in particular avoid unscheduled downtime\footnote{Despite recently publishing books on best-practices for distributed systems design~\cite{book:design:Burns18} and on site reliability engineering~\cite{book:google:SRE16,book:google:SRW18}, since the books were published both the Microsoft and the Google clouds have suffered unscheduled downtime and its related bad publicity.}, poor performance, and the resulting bad reputation.
However, monitoring only reveals what is measurable and measured~\cite{DBLP:journals/cacm/BouwersVD12}, leaving organizations exposed to wicked problems~(defined in Section~\ref{sec:motivation:wicked}) and complex ecosystems~(Section~\ref{sec:motivation:mcs}).

If lackluster design is costly, bad design can be catastrophic. 
Design by committee~\cite{design:laws:Conway68} is known to cause entire projects to fail~\cite[Ch.4]{design:book/Brooks10}, yet many organizations still rely on design by committee done by a central team for technology architecture. 
A particularly bad case of design by committee is when the entire community ignores the needs of the market and society; fiery arguments in this sense appeared in the databases and grid computing communities, around the start of the 2010s.

\vcutS{}
\subsection{The Increasing Need for Good Design} \label{sec:nfd} \label{sec:needfordesign} \label{sec:motivation:needfordesign}
\vcutM{}

\begin{figure}[!t]
  \centering
  \includegraphics[width=0.8\columnwidth]{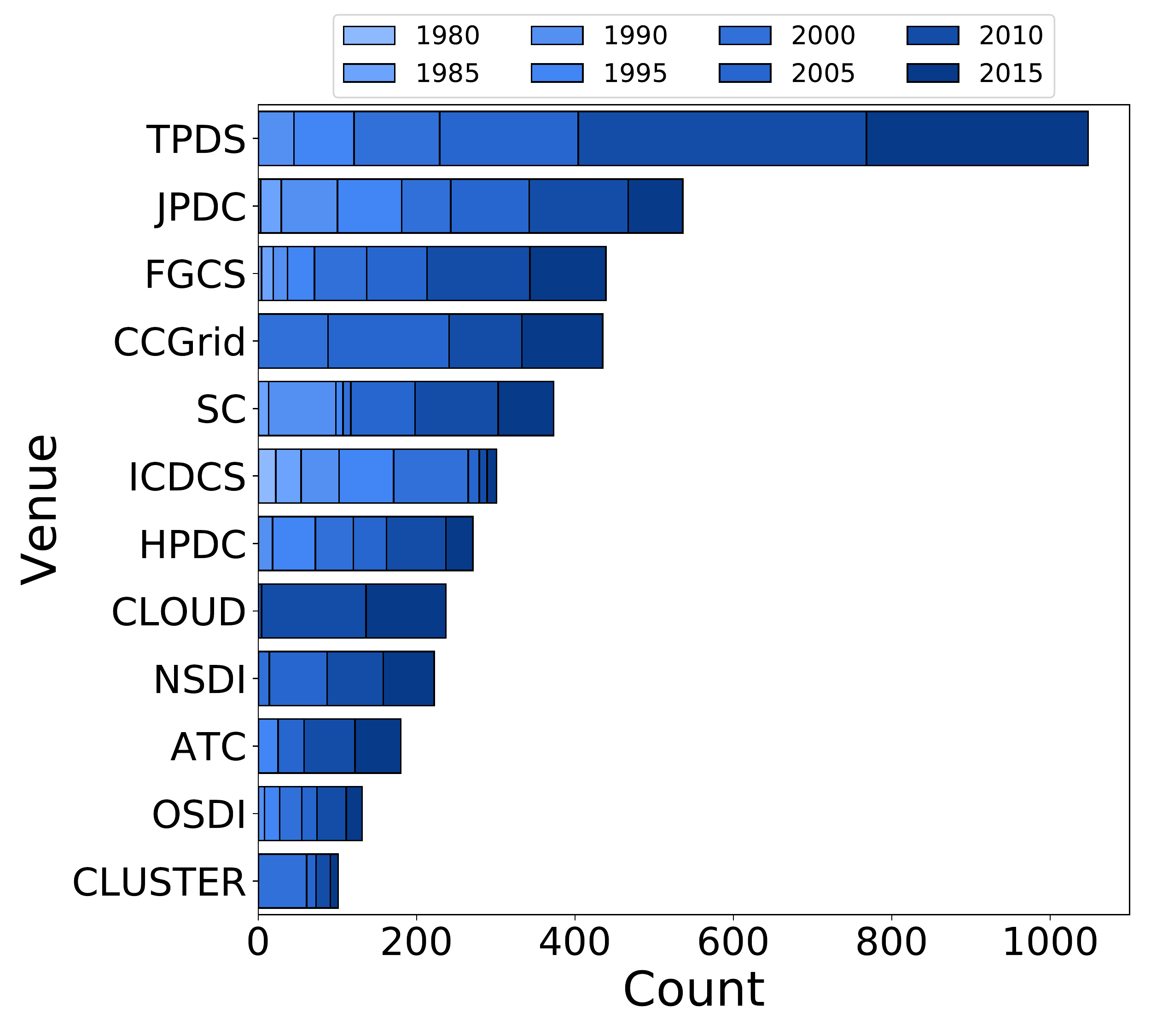}
  \vspace*{-0.5cm}
    \caption{Count of design articles in selected high-quality computer systems venues, since 1980, in 5-year blocks.}
    \vspace*{-0.35cm}
    \label{fig:motivation:fiveyear}
\end{figure}

{\bf Design articles are increasingly present in major distributed systems venues} (Figure~\ref{fig:motivation:fiveyear}).
Complementing the findings related to Figure~\ref{fig:designkeyword}, we ask {\it Is the presence of design articles in top distributed systems venues increasing?} 

We have extracted all design articles appearing in such venues over a period of nearly four decades (from 1980 to 2018), and counted them per venue and per 5-year block.
Figure~\ref{fig:motivation:fiveyear} depicts 
the count 
of design articles in selected systems venues, over contiguous 5-year periods starting with 1980. Some of the venues have started earlier, so for them only censured data is available. The last period depicted in the figure, starting in 2015, is incomplete. 
Many of the venues, including ICDCS, have experienced an increasing accumulation of design articles, with a marked increase in design articles accepted for publication since 2000.

\subsection{The Increasing Need for Good Designers} \label{sec:needfordesigners} \label{sec:motivation:needfordesigners}

We also anticipate
an increasing need for good designers. We identify two main possible sources for good designers: professionals in the field and students about to become such professionals. We analyze here their capabilities, and conclude there is much room for improvement.

\begin{figure}[!t]
  \centering
  \includegraphics[width=\columnwidth]{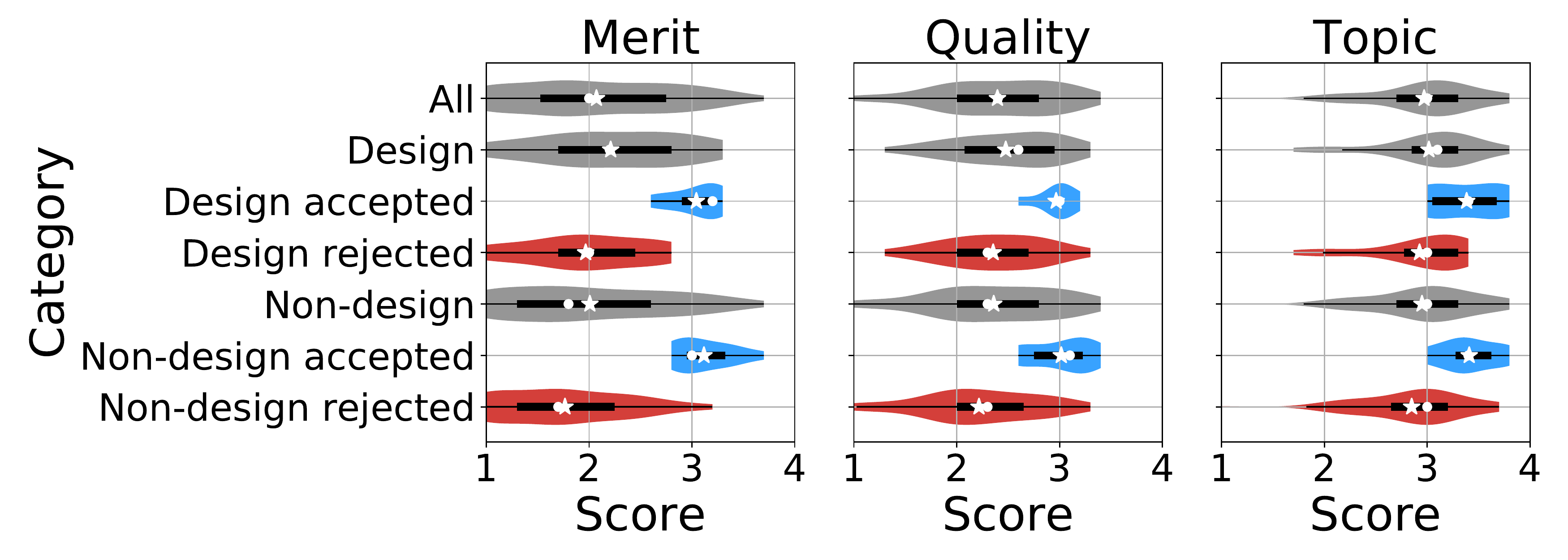}
  \vspace*{-0.75cm}
    \caption{Violin plots of the scores received by articles in a top quality conference in distributed systems, held in the past 5 years.
    Articles are grouped in various categories for the purpose of this analysis.
    Each article is scored for a number of aspects: 
    ({\bf left}) overall merit assessed for each article, ({\bf middle}) quality of contribution, ({\bf right}) match with conference topic. All scores are integers between 1 (lowest) and 4 (highest).
    (Stars depict averages. White dots depict medians. The thick bar denotes the IQR range. Whiskers show 1.5\,$\times$ the range, clipped to actual min and max.)}
    \label{fig:violin-merit}
  \label{fig:violin}
    \vspace*{-0.25cm}
\end{figure}

{\bf Some professionals produce good designs, but still many do not} (Figure~\ref{fig:violin-merit}).
We analyze for a top conference in large-scale distributed systems all the review-results in one year\footnote{We anonymize the venue, but consider it relevant because its held year is after 2014, the venue is a conference, and its ranking is A in CORE18 and green in MSAR14. For comparison, ICDCS has these rankings too.}. For this conference, for each article we have collected whether it is a design article, the final status as accepted for publication in the conference or rejected, and, across the (3+) reviewers, the final scores for (i) the overall quality of the work (the {\it merit}), (ii) the quality of the approach ({\it quality}), and (iii) the fit with the topic of the conference ({\it topic}).
Figure~\ref{fig:violin-merit} depicts the final scores, using distributional (violin) plots. 
For merit, we find that (1) design articles have a slightly better distributional shape over non-design articles, with higher (better) median, mean, and IQR, and more of the distribution around an overall score of 2 or higher.
Across merit and quality, we also find that: (2) a significant percentage of the design articles are not of high quality or high merit (scores significantly below 3). 
Finding (2) is
surprising, because top-tier venues imply self-selection against submitting what the authors themselves consider insufficiently good work; few should merely try out submitting an article to, e.g., ICDCS. This indicates that many professionals still have trouble in both producing 
and self-assessing their designs.

\begin{figure}[!t]
\centering
    \includegraphics[width=\columnwidth]{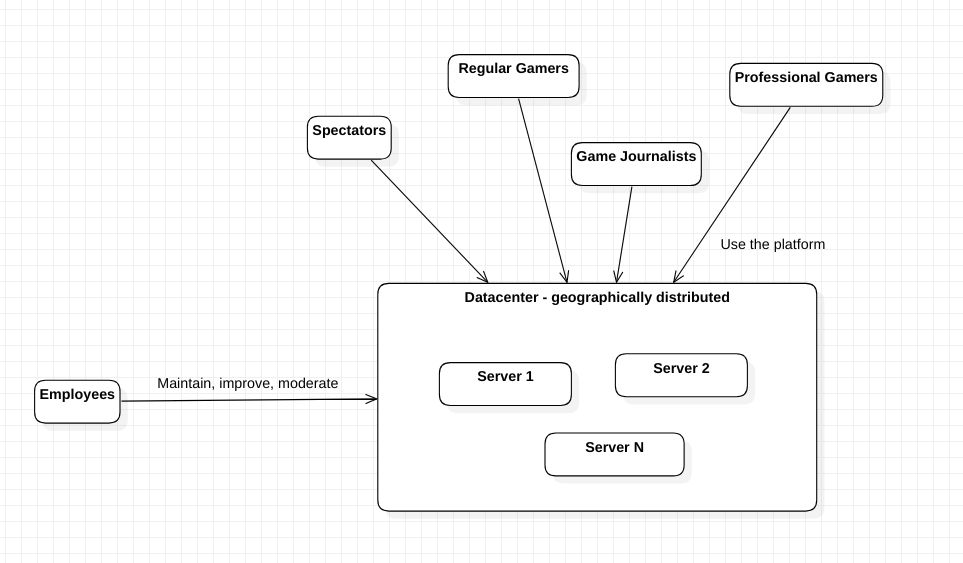}
    \vspace*{-1.0cm}
    \caption{Typical student design, produced early in a graduate distributed systems course at a top university in computer science. The text is difficult to read, as designed by the student.}
    \label{fig:design:student}
    \vspace*{-0.5cm}
\end{figure}

{\bf Graduate students also need training in design thinking and design skills} (Figure~\ref{fig:design:student}).
We analyze here the results obtained from a class of nearly 100 students enrolled in a graduate-level Distributed Systems course\footnote{We anonymize the university, but consider the course relevant because it is large, it took place after 2014, and the university is ranked in the top-150 (in computer science) in both the THE and the QS 2018 World University Rankings (out of nearly 1,000 universities), and in Webometrics of July 18 (out of over 28,000).}; the course seems popular, as the typical class size is around 15 students. We teach in this course not only typical systems concepts from the field, but also concepts and a process for (\ourdomainshort{}) design based on the \AtLarge{} design framework (see Section~\ref{sec:core}). Throughout seven design sessions, students in groups of up to six are tasked to create several designs addressing given problems. 
Figure~\ref{fig:design:student} depicts an early design, attempting to satisfice~\cite[p.27]{book/Simon96} the problem of scalable ecosystems for massivizing an online game~\cite{DBLP:conf/ccgrid/ShenIE13}. The figure represents, to a degree, the common submitted design (across all groups) in the same session---what students know after a Bachelors and some graduate courses, but before learning specifically about design. The figure raises many questions about the quality and even the meaning of the proposed design. Even though it is a simplified and high-level design, it still lacks a believable description for solving (even part of) the problem. For instance, an important missing detail are the interconnections, in the geo-distributed datacenter and between stakeholders. This design also lacks any layering, system packaging, or description of the (sub)components. 
The visual depiction designed by the students is also lacking.

\subsection{New Thoughts on Traditional System Design}
\label{sec:motivation:traditional}
\label{sec:motivation:coevolution}
\label{sec:motivation:wicked}
\label{sec:motivation:illdefined}

System design has gone through successive waves of (shifting) traditional challenges.
The 1950s and 1960s system designers were operating in a world where the core problems seemed structured, and the core design approach could be entirely rational, aiming to {\it optimize} the result~\cite[Part I]{design:book/Brooks10}. {\it Well-structured problems} have several important characteristics~\cite{DBLP:journals/ai/Simon73}: 
(1) a criterion to automatically evaluate the result, 
(2) an unambiguous representation for the goal, and start and intermediate states of the problem, and legal transitions between them,
(3) a clear representation of all domain knowledge,
(4) if interfacing with the natural world, the interaction system-nature can be captured accurately,
(5) the problem itself is tractable.
By the 1970s, it has become apparent that core problems could further be {\it ill-structured}~\cite{DBLP:journals/ai/Simon73}, that is, not have one or several of the characteristics of well-structured problems, or, worse, {\it wicked problems}~\cite{design:RittelW73}, that is, without clear and final formulation, with no universally accepted criteria for success and clearly defined states due to involvement of various stakeholders with competing interests and views, and of various types of hardware and software with various degree of autonomy and limited ability to sense their surroundings.

To address ill-defined and wicked problems, the design community has shifted to {\it satisficing} instead of optimizing designs, and to a process of {\it co-evolving problem-designs}~\cite[Loc.935]{design:book/Brooks10}. 
A cycle of continuous reaction and adaptation triggers the co-evolution: clients change workloads and SLAs, or laws and standards change; in response, system designers evolve, adapt, and decommission parts of the ecosystem; this triggers another round in the cycle. 
Co-evolving problem-desings are typical in systems design~\cite{design:book/Brooks10,DBLP:conf/icdcs/IosupUVAEHTBT18} and pose very significant challenges, in particular because the end-goal is unknowable.
For example, Google's datacenter networking evolved significantly over a decade~\cite{DBLP:journals/cacm/SinghOAAABBDFGK16}, 
as did Google's Spanner for over 5 years~\cite{DBLP:journals/tocs/CorbettDEFFFGGHHHKKLLMMNQRRSSTWW13}.

\subsection{New Challenges in \ourdomainshort{} Design}\label{sec:motivation:content} \label{sec:motivation:mcs}

We identify three major trends and related challenges in distributed systems and ecosystems:

{\bf (C1) New ecosystem life-cycles:}
Whereas in the past many systems were developed and hosted in-house, over the past decade organizations have increasingly shifted operations to (public) cloud computing~\cite{tr:ec14cloud,tr:gartner17cloud}, and thus bought into distributed ecosystems. Consequently, systems and workloads have become much more fragmented than in the past, requiring new approaches for (automatic) decomposition and orchestration. This leads to unexplored design directions in distributed systems, e.g., a strong drive to making them as flexible and composable as possible. 
This further raises many new challenges, e.g., the fundamental challenges of \ourdomainshort{}~\cite[S2.2]{DBLP:conf/icdcs/IosupUVAEHTBT18} are about the lack of: 
(1) operational laws and theories for ecosystems,
(2) comprehensive means to maintain existing ecosystems,
(3) means to explore credible future ecosystem designs,
(4) qualified personnel,
(5) adequate inter-disciplinary tools to assess and control the (unwanted) impact of ecosystems on society.

{\bf (C2) New ecosystem needs and phenomena:}
New design aspects appear when designing entire ecosystems or systems operating in ecosystems. In \ourdomainshort{} systems have many new NFRs, including various forms of elasticity~\cite{DBLP:journals/tompecs/HerbstBKOEKEKBA18}, privacy, interoperability, and operational risk associated with them. 
Ecosystems are {\it super-distributed}~\cite{DBLP:conf/icdcs/IosupUVAEHTBT18}: they are recursively distributed, with their constituents often being distributed (eco)systems; yet, FRs and NFRs in distributed systems are not known to be directly composable across ecosystems.
Various {\it dynamic phenomena} appear in distributed ecosystems, seemingly unique situations that do not fit the patterns expected from current theory and practice; for example, {\it vicissitude}~\cite{DBLP:conf/ccgrid/GhitCHHEI14} is a class of phenomena where several known bottlenecks appear seemingly at random in various parts of the system, {\it performance variability} is common in clouds~\cite{DBLP:conf/ccgrid/IosupYE11}, datacenter networks~\cite{DBLP:conf/sigcomm/BallaniCKR11}, and big data operations~\cite{DBLP:conf/wosp/UtaO18}, and ecosystem owners spar with each other (e.g., in Jan 2019, Apple denied Facebook and Google access to its APIs, Unity changed their Terms-of-Service and thus locked out small developers like SpatialOS).

{\bf (C3) New ecosystems, old parts:}
The evolution of distributed systems technology has generated many useful parts that are commonly used in today's ecosystems, 
from simple mechanisms (e.g., caching, scheduling), 
to protocols (e.g., for multi-site data transfer) and policies (e.g., for autoscaling), 
to relatively simple systems (e.g., BitTorrent for file-sharing), 
to commonly used architectures (e.g., for web applications, for big data processing). 
A large amount of legacy applications, using various generations of technology, still operate. 
Yet, this legacy technology and applications were not designed for the new ecosystems, for example, they are not cloud-native. Fully replacing them could be prohibitively expensive in the short-term, which means \ourdomainshort{} designers must innovate to keep them operational, efficiently.

%% file: content/03-atlarge-framework.tex
\section{The \AtLarge{} Design Framework for \ourdomainshort{}}
\label{sec:core}

In this section, we summarize the current theories about design as an activity, then focus on the \AtLarge{} design framework. 
We give its central premise, explain the focus and main concerns, and focus on its key methods for design-space exploration, problem-finding, problem-solving, and reporting to the community and society.
Overall, the key contribution of this framework is that it combines current theories about design thinking (Section~\ref{sec:core:designerly}) with \ourdomainshort{}-focused design processes (Sections~\ref{sec:core:overview}--\ref{sec:core:reporting}).

\vcutS{}
\subsection{Designerly Ways of Thinking}
\label{sec:core:designerly}
\vcutM{}

\begin{figure}[!t]
    \centering
    \includegraphics[width=\columnwidth]{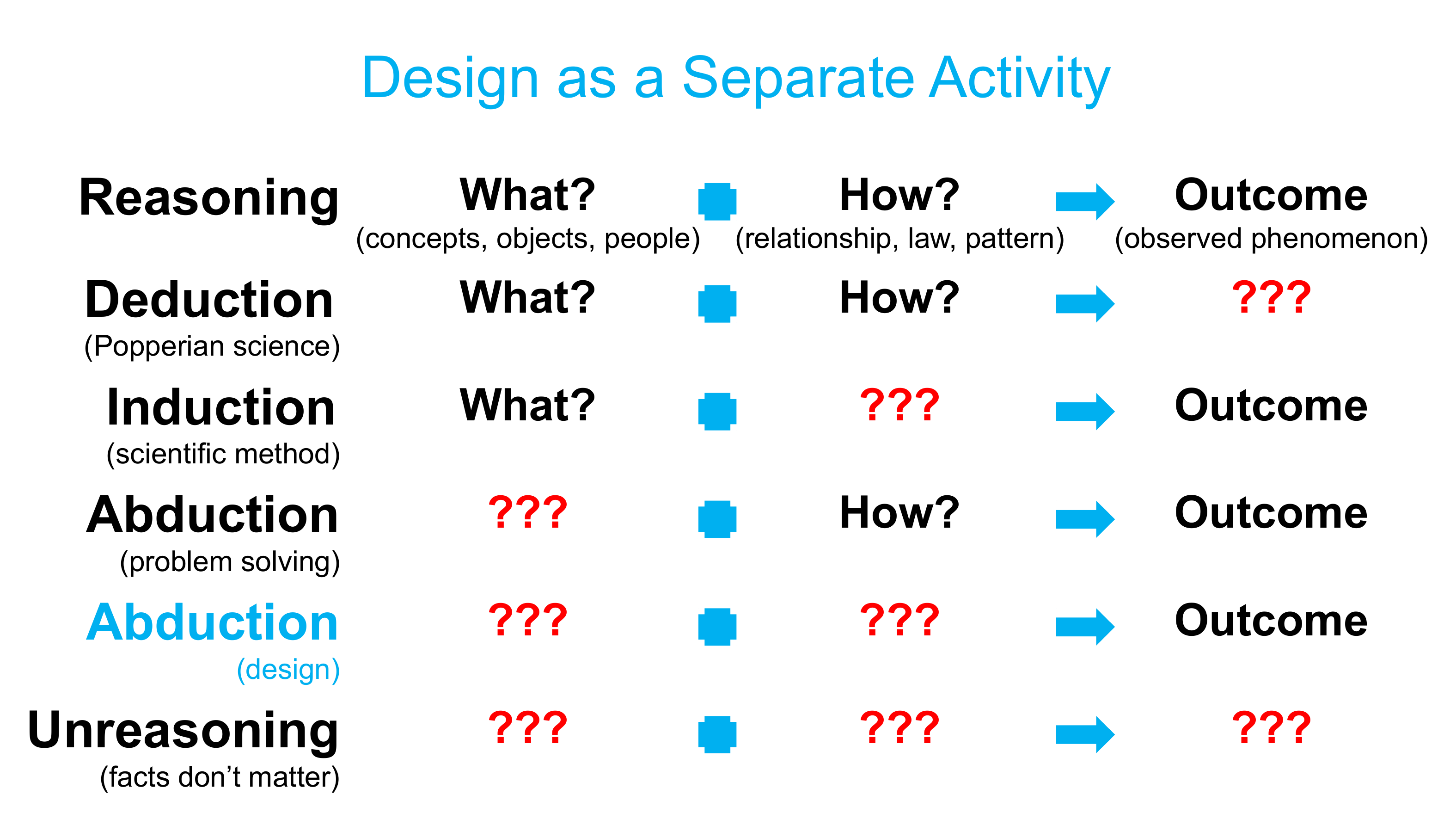}
    \vspace*{-0.75cm}
    \caption{Dorst considers design an unique intellectual activity, based on abduction~\cite[p.13]{book:design:Dorst17}. In particular, design is not science or engineering.}
    \label{fig:design:unique}
\end{figure}

{\bf Design, from engineering component to independence\footnote{Computer and software engineering have traversed a similar process until emancipation in the mid-1960s~\cite[Part III]{book:philosophy:Tedre15}, when detaching from mathematics. Interestingly, mathematics had to follow a similar process, to detach from philosophy; an important part of this process Hilbert's program~\cite{book:math:Franks09}.}:}
Ever since the introduction of the concept of ``designerly ways of thinking'', in the 1990s~\cite[p.68, concept by Cross]{design:book/Lawson05}, and possibly also earlier, the modern design community has held as a theoretical principle that design is based on specific, idiosyncratic ways of thinking. 
In 2017, Dorst described a theoretical model for reasoning~\cite[p.13]{book:design:Dorst17} that includes design thinking, in which the reasoning universe consists 
of specific concepts (e.g., real people, software objects), which represent the ``What?'' of the problem to solve; 
of relationships between the concepts (e.g., laws of nature, principles of hardware operation, software patterns), which represent the ``How?'';
and of an outcome that combines the concepts and the relationships (e.g., into a real-world system, into an observable phenomenon).

Figure~\ref{fig:design:unique} depicts the Dorst reasoning model.
In this model, {\it deduction} proceeds from given concepts and relationships, and reasons toward an outcome that can be observed (and, thus, testing the deduction); for example, given a Turing machine and a deterministic algorithm designed for it (and its input), we can deduce its outcome. {\it Induction} follows another classical model from science. 
Abduction for problem solving ({\it normal abduction} in Dorst's model) matches well the {\it software engineering} experience---given the architecture of a software system, determine the best software-design patterns, and the other software engineering concepts and objects, to realize the system that would act as predicted at design-time.
{\it Unreasoning}, which we add to the Dorst model, simply states an extreme of reasoning where any concept, relationship, and outcome can be put together, for example, by an organization for which facts do not matter (one of ``alternative facts'').

{\bf Design abduction:}
In contrast to the other reasoning approaches in Figure~\ref{fig:design:unique}, {\it design abduction} begins with a desirable outcome, and the problem becomes one of finding the concepts and their relationships that lead to the outcome. 
Of course, an intractable or even infinite number of possible concepts and relationships can exist to consider, which is what makes the design problem rarely amenable to normal abduction (and normal engineering).
This does not mean that design abduction must be purely creative, without process.

\begin{table}[!t]\centering
\ra{1.1}
\begin{tabular}{@{}lll@{}} \toprule
Who? & Stakeholders       &  designers, scientists, engineers, \\
     &                    &  students, society \\
\midrule
What? & Central Paradigm  &  design, different \\
      &                   &  from science and engineering \\
      & Focus             &  ecosystems, systems within\\
      &                   &  structure, organization, dynamics  \\
      & Concerns          &  functional and non-functional  \\
      &                   &  properties; phenomena, evolution \\
\midrule
How?  & Design Thinking   &  abductive thinking, processes,  \\
      &                   &  co-evolving problem-solution  \\
      & Exploration       &  design space, process to explore\\
      & Problem-finding   &  structured, ill-defined, wicked\\
	  & Problem-solving   &  pragmatic, innovative, ethical\\
	  & Reporting         &  articles, software, data \\
\bottomrule
\phantom{abc}
\end{tabular}
\vcutM{}
\caption{An overview of the \AtLarge{} design framework.}
\label{tab:domain:design}
\vcutL{}
\end{table}

\subsection{Overview}
\label{sec:core:overview}

We give an overview of the \AtLarge{} design framework and summarize its key properties in Table~\ref{tab:domain:design}: Who? What? How? are the questions addressed in this section.

{\bf Who? Stakeholders:} The primary stakeholder of \ourdomainshort{} design is the society; this is because designs in this field can have an unusually large impact, for a direct product of computing. The \AtLarge{} design framework considers explicitly that designers fulfill a separate role from scientists and engineers, and, consequently, that students require explicit training in design.

{\bf What? The Central Premise: design is unique among intellectual activities.}
Like Cross~\cite{design:book/Cross11}, Dorst~\cite{book:design:Dorst17}, and Parsons~\cite{book:design:Parsons15}, the \AtLarge{} framework considers design an unique intellectual activity, essentially different from science and engineering. This does not mean that scientists and engineers cannot design---theory and practice indicate all people can and do design naturally~\cite[Loc.275, theory by Victor Papanek]{book:design:Parsons15}---, but doing so proficiently and efficiently still requires professional expertise, much like engineering and science. 

{\bf What? The Main Focus and Concerns: support for \ourdomainshort{} design.} 
This requires focusing on both the traditional challenges raised by system designs~(see Section~\ref{sec:motivation:traditional}) and the new challenges raised by \ourdomainshort{}~(see Section~\ref{sec:motivation:mcs}).
Two traditional problems of design are to identify the design space and to explore it efficiently; how to do so for \ourdomainshort{} designs is an open challenge.
Among the \ourdomainshort{}-specific aspects, the \AtLarge{} design framework considers explicitly, for every problem:
the architecture of ecosystems and of systems operating in ecosystems;
the structure, organization, and dynamics of ecosystems;
functional and non-functional properties and their expression as implicit (that is, designer-given) or explicit (that is, client-given) SLAs;
and known aspects of ecosystem phenomena, emergence, evolution.

\begin{figure}[!t]
    \centering
    \includegraphics[width=\columnwidth]{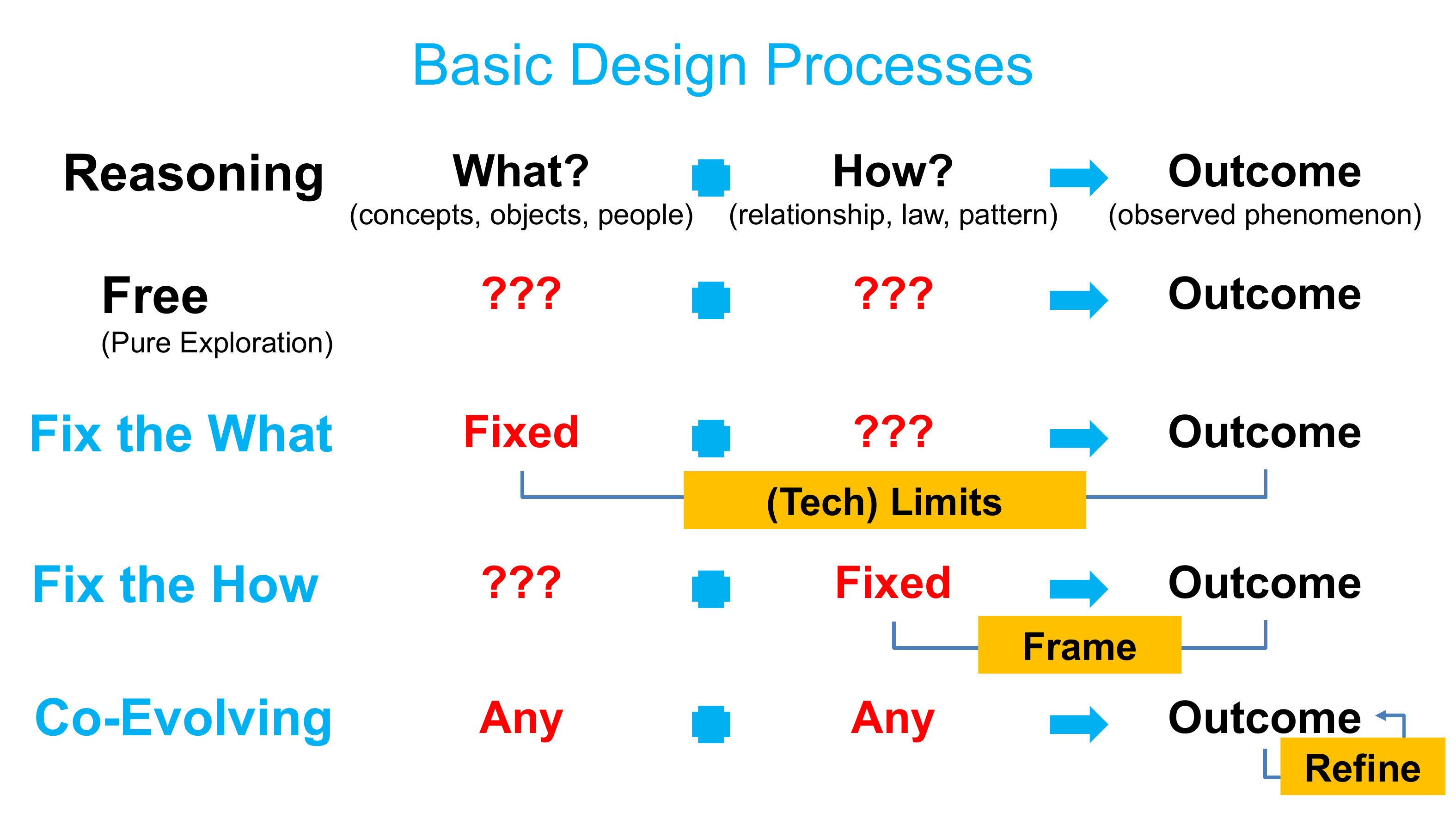}
    \vspace*{-0.75cm}
    \caption{Basic design processes address how to explore the design space. They range from free to fixed to co-evolving exploration.}
    \label{fig:design:processes}
\end{figure}

{\bf How? Designerly Thinking:} Derived from its central premise, the \AtLarge{} design framework considers designerly thinking as an essential ability of its practitioners. Among its core elements, this ability includes understanding, conducting, and managing design as co-evolving problem-solutions. Additional reasoning and practical skills related to science and engineering are also welcome. 

{\bf How? Key Processes:} 
Although in practice design is still largely an unstructured process, and attempts to impose a rigid structure cause negative reactions~\cite{design:book/Brooks10} and even opposition in software engineering practice\footnote{The agile manifesto, \url{https://agilemanifesto.org/}}, the \AtLarge{} design framework holds that there still is room for (flexible) process for design.
Key to good design, the framework proposes not rigid steps, but a small number of flexible methods and processes for: design space exploration (in Section~\ref{sec:core:exploration}), problem-finding (in Section~\ref{sec:core:findproblem}), a basic cycle for problem-solving (in Section~\ref{sec:core:solveproblem}), and for making the results available beyond the design team (in Section~\ref{sec:core:reporting}).

\vcutS{}
\subsection{Free to Co-Evolving Design Exploration}\label{sec:core:process}\label{sec:core:exploration}
\vcutM{}

{\bf A general, flexible approach to design space exploration for \ourdomainshort{}:}
Figure~\ref{fig:design:processes} depicts several processes for design exploration.
Following the Dorst design framework, the design abduction could be conducted freely, as pure exploration: the designer considers concepts and relationships at will, guided by own intuition and shared community expertise. Although this approach can result in radically new designs, its likelihood of success is limited by the scale of the design space. 
In contrast, the \AtLarge{} design process considers three other, more structured approaches for design space exploration. All three consider that there is a process for finding good problems, for example, the process described in Section~\ref{sec:core:findproblem}.
The {\it Fix the What} and {\it Fix the How} processes explore the same trade-off: they aim to improve the likelihood of obtaining satisficing designs by diminishing the likelihood that the design will be radically innovative. They both do this by limiting the options available to explore. The former does this by fixing the concepts at play and in particular the technology the designer can use; the latter, by fixing the kinds of relationships available to the user (``(re-)framing'' in traditional design~\cite[p.14]{book:design:Dorst17}). 

The third process, {\it co-evolving}, focuses on iterating designs by changing the problem itself, and further allows using any of the other exploration processes for solving the problem in the current iteration. The staple of this process is the co-evolving problem-solution, with which it can explore a potentially unlimited design space while having a satisficing solution available at each iteration (after the first iteration). 

\begin{figure}[!t]
    \centering
    \includegraphics[width=\columnwidth]{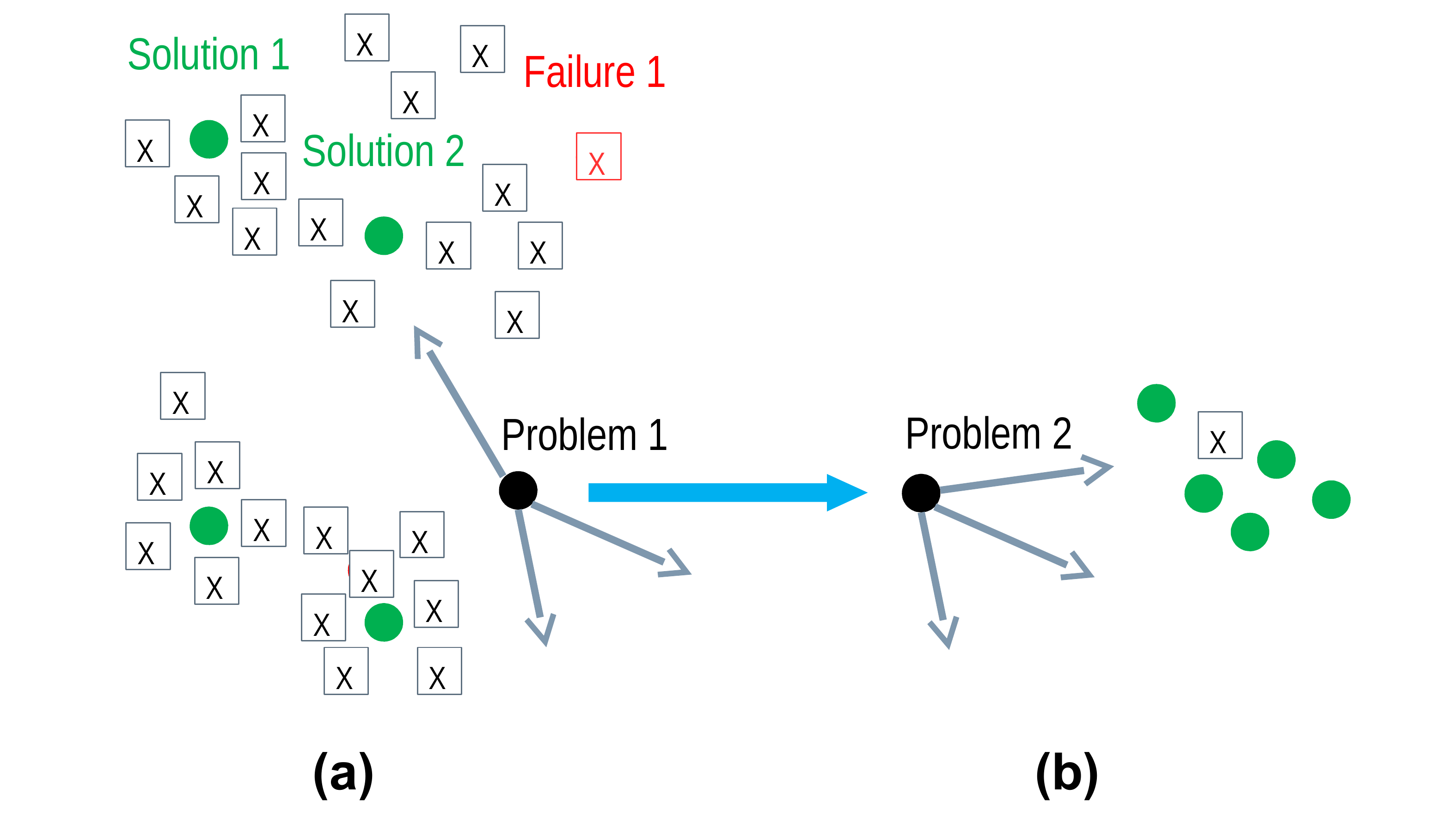}    \vspace*{-0.75cm}
    \caption{Design exploration for \ourdomainshort{}, through an example. Dark circles are problems. Light (green) circles are successful designs. Boxes marked with an ``X'' are design attempts that result in failure.}
    \label{fig:design:exploration}
\end{figure}

{\bf How does co-evolving design space exploration work, in practice?}
Figure~\ref{fig:design:exploration} depicts an abstract but realistic example of co-evolving design. 
The Design Team (DT) is trying to create a pragmatic, innovative design. 
DT starts with a problem (Problem 1 in Figure~\ref{fig:design:exploration}~(a)).
DT creates a design for it, which satisfices or even optimizes the problem (Solution 1). It is not too sophisticated, so DT agrees they could do better.
DT tries to do better, and fail (Failure 1). 
DT learns from it, and produce a new design (Solution 2).
Iterating through their design cycle, DT keeps traversing the design space, exploring several dimensions concurrently, and find after much struggle (and failure) a couple more solutions. However, at this point DT concludes it is too difficult and/or costly to keep exploring.
DT has learned enough in the process of design, and possibly with help from their community and clients, and area ready to evolve the problem (Problem 2 in Figure~\ref{fig:design:exploration}~(b)); for example, DT could focus the design on a new ecosystem, replacing the old ecosystem that proved to be too limited for DT to solve the problem. (This does not mean the old ecosystem is not good for other design teams or for other design problems.)
It turns out that, for this new problem, DT can find many new solutions relatively easily. The process is successful, and promises more success for the future.

\subsection{Problem-Finding Process for \ourdomainshort{} Design}\label{sec:core:findproblem}\label{sec:core:knowledge}

{\bf Approach:} It would be presumptuous to claim there exists a process for finding all the problems \ourdomainshort{} designers can solve. Instead, inspired by how conferences in the field use Calls for Papers to steer the authors, the \AtLarge{} design framework 
aims to focus the designer by proposing a set of {\it problem archetypes} (topics). The community could help expand and refine this set in the future. 
This approach seems highly successful in focusing designers---Figure~\ref{fig:violin}~(right) indicates 
the designs submitted for evaluation match closely the topics proposed by the conference's community, as proposed by the Program Chairs. 
Although none of the concepts used in the framework is new, synthesizing these aspects into a catalog, as we do in this work, is novel for the field and seems valuable (see Section~\ref{sec:exp}).

{\bf What kinds of problems?} Derived from Section~\ref{sec:motivation:mcs}, 
the \AtLarge{} design framework proposes to focus on: 
(P1) problems in ecosystem life-cycle, including for new and emerging processes and services, and for new and emerging ecosystems;
(P2) problems related to new and emerging needs of ecosystem-clients and -operators, addressing newly discovered, emerging, and recurring phenomena, and harnessing new technology (a special kind of phenomenon);
(P3) problems related to leveraging and maintaining legacy components.
Besides problems that lead to creating new technology, 
(P4) inspired by natural sciences, where understanding the morphology of natural ecosystems is valued, problems related to understanding how new and emerging technology actually works in practice or when placed in ecosystems, and what new phenomena appear related to ecosystem-operation;
(P5) inspired by mathematics, where creating new abstractions can be important regardless of application, problems related to previously unexplored parts of the design space.

{\bf How to identify meaningful problems?}
Also here, the \AtLarge{} design process tries to select from known approaches to identify problems. For addressing problems of types (P1)--(P3), the designer could try to collect and adapt problems from various sources: 
(S1) (peer-reviewed) qualitative and quantitative studies conducted on ecosystems and on systems within them;
(S2) discussion with experts, own analysis of best-practices including reading of technical reports, tech blogs, and best-practice books;
(S3) own thought and lab experiments concerning the key technology trends, known technical and other limitations, etc.

For P4, the designer could follow a process matching (empirical) science, but focusing on systems, leveraging the scientific process as finder of phenomena to be harnessed. This could include understanding how systems work through collection and analysis of data archives, where the data represents workloads (e.g., structure of jobs, job life-cycle events such as arrivals, migrations, and cancellations) and operations (e.g., utilization of specific components, (un)availability events). Here, an important set of problems relate to collecting meaningful data: the construction of the observation or measurement instrument, the design of a meaningful data-collection protocol, etc. Currently, these problems seem largely ignored in our field, leading to a dearth of meaningful data for experiments and, possibly, for discovering real problems.

For P5, derived from the notions of views in software engineering~\cite{Rozanski2005} and of morphological analysis in sciences~\cite{Zwicky48}, the designer could identify unoccupied niches and formulate the problem of exploring them, driven by curiosity.

\subsection{Problem-Solving Process for \ourdomainshort{} Design}\label{sec:core:solveproblem}\label{sec:core:design}

{\bf Approach:} Similar to problem-finding, problem-solving is too diverse to capture in any single process; moreover, stage-based processes can raise resistance from practitioners as too constraining~\cite{design:book/Lawson05}.
The \AtLarge{} design framework aims to balance the pragmatic need to have a process with clear stages, which allows teams to synchronize about and during the process of design, and the need for innovation that is based on the flexibility to not stifle creativity. 
To this end, the framework includes an iterative process focusing on creative tasks, which in particular allows its practitioners to skip any step at each iteration. 
Unlike typical processes in the field, which focus either on hardware design~\cite{book:design:BlaauwB97,book:compsys:HennessyP17} or on software design~\cite{book:design:Abbott15,book:design:Bass15}, or on higher-level processes on keeping the team agile~\cite{DBLP:journals/csur/RamsinP08}, the \AtLarge{} problem-solving process focuses first on system-level concerns. Pragmatically, this means it considers first the concepts, components, and challenges specific to \ourdomainshort{}. 

To manage the complexity of the problem of designing distributed systems and ecosystems, the \AtLarge{} problem-solving process includes two core elements: (1) a {\it Basic Design Cycle~(BDC)}, which is a general process for solving problems, and (2) an {\it Overall Process} that combines several BDCs into a structure for decomposing and solving \ourdomainshort{} design-problems.
We have detailed our problem-solving process elsewhere~\cite{conf/hpdc/DesignProcess19}, and only summarize it here.

{\bf The BDC is the core loop:} The BDC process aims to solve any generic design problem through a structured process consisting of the following elements: (1) Formulate requirements, (2) Understand alternatives, (3) Bootstrap the creative process, (4) High-level and low-level design, (5) Implementation of mathematical analysis code, of simulators, of prototypes, etc. (6) Conceptual analysis of the design, (7) Experimental analysis of the design, (8) Result summarizing and dissemination. This approach is by design: it matches many classic design processes, and is recognizable to designers and engineers in the distributed systems field, yet each element includes key innovations~\cite[Table 1]{conf/hpdc/DesignProcess19}.

{\bf The Overall Process~(OP)} is executed iteratively. It operates as an BDC and, hierarchically, its more complex {\it design stages} can also operate as BDCs. This design of the OP allows the designers to partition into manageable parts the inherently complex process of solving the  problems typical of \ourdomainshort{} design, e.g., formulating requirements, creating believable designs\footnote{That the result is {\it believable} is the core of the epistemological problem of design~\cite[Loc.972]{book:design:Parsons15}. It is even more so for \ourdomainshort{}-designs, because that such designs are unlikely to be analyzed experimentally to the full extent of their intended application; in other words, many designs will at best be shown as believable, through narrow laboratory experiments.}.
The hierarchical nature of the OP further facilitates learning the process by practitioners: once a practitioner has learned the BDC, they can apply it several times in the OP. 

The OP has one more important feature: in each iteration, each of its stages can be skipped as needed. By not forcing \theDesigner{} to traverse unnecessary elements, the OP allows each iteration to be tailored to the remaining parts of the problem to be solved, and to the remaining time and other resources. We conjecture this can lead to designerly thinking~(see Section~\ref{sec:core:designerly}).

\begin{figure}[!t]
    \centering
    \includegraphics[width=0.6\columnwidth]{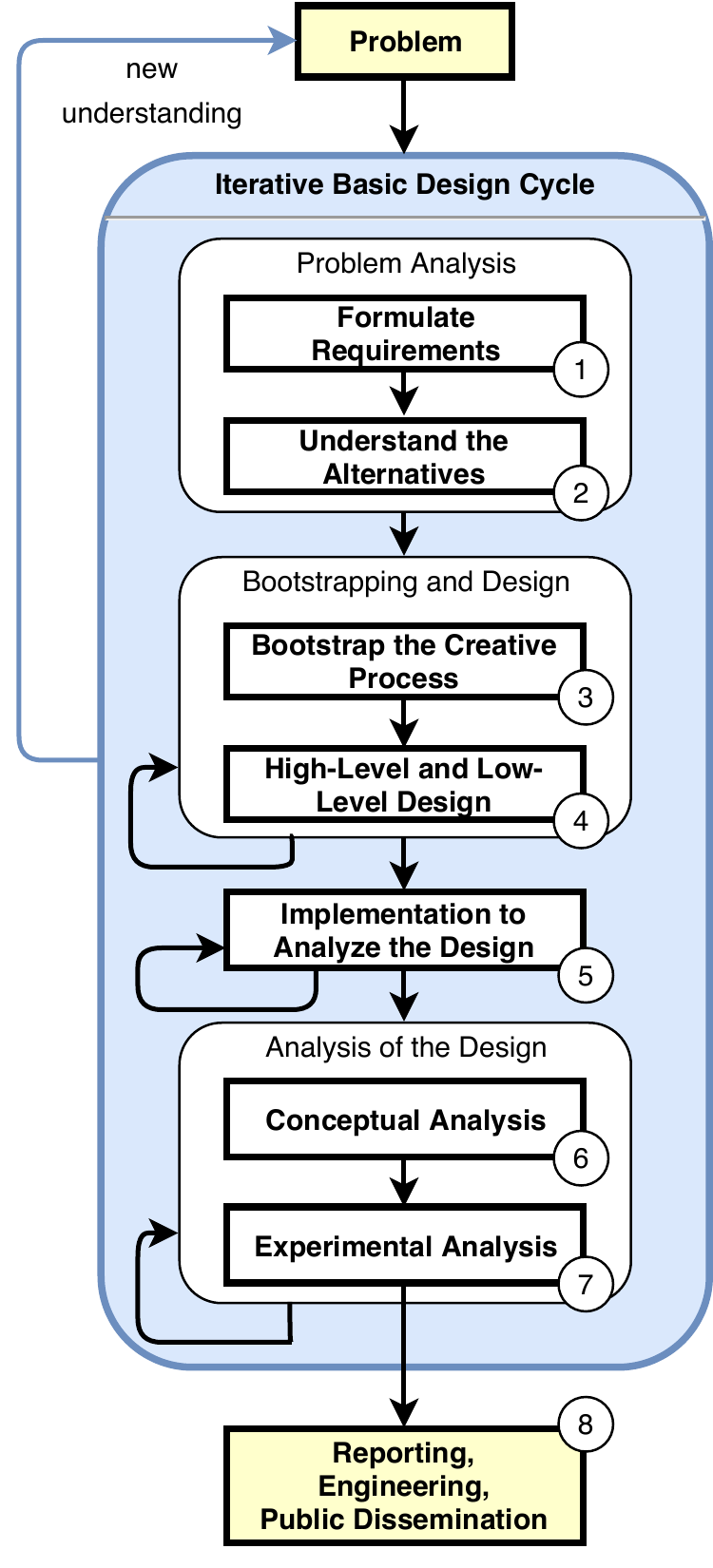}
    \vspace*{-0.25cm}
    \caption{The \AtLarge{} design process.}
    \label{fig:design:problemsolving}
    \vspace*{-0.5cm}
\end{figure}

{\bf The OP elements:}
Figure~\ref{fig:design:problemsolving} depicts the OP. Given a design problem, its BDC spans elements 1--8, with various groupings allowing for finer-grained iteration. 
In the overall BDC process, elements~(5) and~(7), which can include various types of prototype implementation and of experimentation, respectively, can be complex. 
When this complexity occurs, the designers need to expand them each into one BDC. 
Similarly, Element~(8), on reporting, engineering, and public dissemination, can further expand into separate BDC processes for publishing articles, free open-access software (FOSS), and FAIR~\cite{data:FAIR16} or free open-access data (FOAD); we explain this element in Section~\ref{sec:core:reporting}.

{\bf Stopping criteria:} As any {\it iterative} process, BDC stops when meeting a predefined set of: 
(1) finding a single answer that {\it satisfices}~\cite[p.27]{book/Simon96}, that is, gives solutions that are ``good enough'', or, where possible, {\it optimizes};
(2) finding a few answers, forming a portfolio to allow a human reviewer (e.g., a client) to quickly select one; 
(3) finding many answers, forming a {\it systematic design}, that allows an expert reviewer or system to select one;
(4) finding all answers, resulting in {\it design space exhaustion} and allowing experts across the community to discuss or select results;
(5) running out of time or other resources (e.g., funding).

{\bf BDC can, but does not guarantee success:} Because it admits the stopping criterion 5, the BDC does not guarantee a result. 
In our experience so far, following the OP process has a good probability of success, making pragmatic and innovative designs likely within the time- and resource-budget. 
We present experimental evidence for this in Section~\ref{sec:exp}.

\subsection{Dissemination Processes for \ourdomainshort{} Design}
\label{sec:core:reporting}

The \AtLarge{} design framework also considers various forms of dissemination typical for \ourdomainshort{}, related to reports, software, and data. 
Each of these means of dissemination is based on some form of design; for example, designing the reports to be published as peer-reviewed articles. 
Thus, for each, the framework proposes design-based processes; in essence, smaller versions of the framework itself, and in particular of the BDC (see Section~\ref{sec:core:solveproblem}). 

The reason for this is similar to the reason to use more structured design processes for \ourdomainshort{}: it should increase the likelihood of good designs. 
Although 
the dissemination of reports, software, and data can be achieved through much intuition, expertise, and by following best-practices in the respective fields (e.g., collaborative editing using a tool such as Overleaf; collaborative FOSS development using CI/CD tools such as Travis CI and customized solutions~\cite{DBLP:conf/msr/WidderHKV18}; and sharing code and data on archives such as GitHub and Zenodo, respectively), in practice many of these designs are poor (see Section~\ref{sec:motivation:needfordesigners}).

%% file: content/04-design-principles.tex
\section{Design Principles of \ourdomainshort}
\label{sec:core:principles}
\vspsmall

\begin{table}[!t]\centering
\ra{1.1}
\begin{tabular}{@{}lrl@{}} \toprule
& \multicolumn{2}{l}{Principle} \\ \cmidrule(r){2-3}
Type (Section) & Index & Key aspects \\ 
\midrule
Highest ($\S$\ref{sec:core:highestprinciple}) & P\ref{pr:design}      &  design of design\\
\midrule
Systems & P\ref{pr:ecosystems}&  age of distributed ecosystems \\
($\S$\ref{sec:core:system})  & P\ref{pr:nfr}           &  NFRs, phenomena\\
 & P\ref{pr:rms}           &  RM\&S, self-awareness\\
\midrule
 Peopleware & P\ref{pr:edu} &  education in design\\
 ($\S$\ref{sec:core:peopleware}) & P\ref{pr:community}&  pragmatic, innovative, ethical \\
\midrule
Methodology & P\ref{pr:science}&  design science, practice, culture\\
($\S$\ref{sec:core:methodology}) & P\ref{pr:evolution}     &  evolution and emergence\\
\bottomrule
\phantom{abc}\\
\end{tabular}
\vspace*{-0.25cm}
\caption{The key principles of \ourdomainshort{} design.} 
\label{tab:principles}
\vspace*{-0.5cm}
\end{table}

We introduce in this section a set of core principles for \ourdomainshort{} design. 
Table~\ref{tab:principles} summarizes the principles.

\subsection{Highest Principle}
\label{sec:core:highestprinciple}

\principle{pr:design}{{\bf Design needs design.}} 
We have argued for this principle in Section~\ref{sec:motivation}.
The highest design principle holds that \ourdomainshort{} design must be designed, not left only to intuition and selective experience.

\subsection{Systems Principles}
\label{sec:core:system}
\label{sec:core:organization}
\label{sec:core:characteristics}

\principle{pr:ecosystems}{{\bf This is the Age of Distributed Ecosystems.}}
As stated in Section~\ref{sec:motivation:mcs}, the evolution of distributed systems into ecosystems led to important new problems and solutions. This principle argues for an approach to design where the designer is constantly aware of this fact.

\principle{pr:nfr}{{\bf Dynamic non-functional properties and phenomena are first-class concerns.}}
\principle{pr:rms}{{\bf Resource Management and Scheduling, and its interplay with various sources of information to achieve local and global Self-Awareness, are key concerns.}}
Principles~\refpr{pr:nfr}{} and~\refpr{pr:rms}{} are consequences of \ourdomainshort{} problems always including dynamic and emergent elements. Good \ourdomainshort{} designs must consider complex SLAs, emergent phenomena, information-rich decision-making, etc.

\subsection{Peopleware Principles}
\label{sec:core:peopleware}

Popular distributed ecosystems service hundreds of millions daily. It is not uncommon for a typical service to call into execution hundreds of hidden systems. This combination of high complexity and responsibility puts pressure on the human resources---the peopleware.

Inspired by the software industry's struggle to manage and develop its human resources, we explicitly set principles about peopleware.

\principle{pr:edu}{{\bf Education practices for \ourdomainshort{} must ensure the competence and integrity needed for experimenting, creating, and operating ecosystems.}} 
Because the complexity and responsibility of the job has increased considerably over the past couple of decades, high-quality design education should become a core principle of \ourdomainshort{}.
With proper training, the community will remain able to produce designs significantly better than early, student-like attempts (see Figure~\ref{fig:design:student}), and avoid a culture of hacking that does not work long-term. 
Education on the ethics of design is also a must, if the community is to avoid even the most basic traps, such as engendering bias and disregarding privacy.

\principle{pr:community}{{\bf Design communities can foster and curate pragmatic, innovative, and ethical design practices.}} 
The community is already structured to foster and curate designs (see Section~\ref{sec:motivation:needfordesign}). This principle extends this structure to include shared tools and environments for developing and evolving designs: shared datasets and benchmarks, testing infrastructure available to many, common repositories of and documents about operational patterns, online virtual laboratories for global coursework and training, etc. These are elements that greatly facilitate design, and make it pragmatic by linking  academia and industry. The community is also best-equipped to understand and explain the ethics of the field, and further to handle ethical risks.

\subsection{Methodological Principles}
\label{sec:core:methodology}

\principle{pr:science}{{\bf We understand and create together a science, practice, and culture of \ourdomainshort{} design.}} 
So far, design has not been treated as a scientific subject in the field of distributed systems. However, design should become such a subject, because it meets the requirements explained by Denning~\cite[p.32]{DBLP:journals/cacm/Denning13a}: 
(i) \ourdomainshort{} design is a pervasive phenomenon, which we try to understand, use, and control; 
(ii) both artificial and natural processes are at play (designs lead to real-world artifacts); 
(iii) we aim to gain meaningful and non-trivial understanding of the phenomenon; 
(iv) we aim to make our findings reproducible, so that good designs become more likely, consequence of falsifiable theories and models; 
etc. 

\principle{pr:evolution}{{\bf We are aware of the history and evolution of \ourdomainshort{} designs, key debates, and evolving patterns.}} 
Unlike other exact results in distributed systems, design is prescriptive, and often discursive. This makes it subject to debate and interdisciplinary expertise. To improve design, we need to make use also of the key instruments of empirical research, including exploring the history of the field, surveying the expert view, understanding the key debates and their ongoing resolution (as Tedre does for the whole field of computer science~\cite{book:philosophy:Tedre15}), etc.

%% file: content/05-design-challenges.tex
\section{Ten Challenges for \ourdomainshort{} Design}
\label{sec:challenges}

Many challenges must be overcome before the principles in Section~\ref{sec:core:principles} can give us a solid basis for design.
Known challenges begin with making the highest principle, of  \ourdomainshort{} design being based on a design rather than on intuition, a reality. Challenges appear also related to systems, peopleware, and methodological aspects. We give in the following a non-exhaustive list of ten challenges for \ourdomainshort{} design.

\begin{table}[!t]\centering
\ra{1.1}
\begin{tabular}{@{}lrll@{}} \toprule
& \multicolumn{3}{c}{Challenge} \\ \cmidrule(r){2-4}
Type (Sec.) & Index & Key aspects & Pr.\\ 
\midrule
Highest & C\ref{ch:design}&  Design of design & P\ref{pr:design}\\
Principle & C\ref{ch:gooddesign}&  What is good design? & P\ref{pr:design}\\
($\S$\ref{sec:ch:highest}) & C\ref{ch:exploration}&  Design space exploration & P\ref{pr:design}\\
\midrule
Systems &   C\ref{ch:ecosystems} & Design for ecosystems & P\ref{pr:ecosystems}\\
($\S$\ref{sec:ch:system}) &  C\ref{ch:catalog}           &  Catalog for \ourdomainshort{} design & P\ref{pr:nfr}--\ref{pr:rms}\\
\midrule
Peopleware  & C\ref{ch:curriculum}    &  Education, curriculum & P\ref{pr:edu}\\
($\S$\ref{sec:ch:peopleware}) & C\ref{ch:engagement}    &  Community engagement & P\ref{pr:community}\\
\midrule
Methodology & C\ref{ch:documenting}& Documenting designs & P\ref{pr:edu}--\ref{pr:science}\\
($\S$\ref{sec:ch:methodology})    & C\ref{ch:ethno}&  Design in practice & P\ref{pr:science}\\
 & C\ref{ch:orgsim}        &  Organizational similarity & P\ref{pr:science}\\
\bottomrule
\phantom{abc}\\
\end{tabular}
\vspace*{-0.25cm}
\caption{A shortlist of the challenges raised by \ourdomainshort{}.}
\label{tab:challenges}
\vspace*{-0.5cm}
\end{table}

\subsection{Challenges Related to the Highest Principle}
\label{sec:challenges:highestprinciple}
\label{sec:ch:highest}

\challenge{ch:design}{\refpr{pr:design}}{The design of design. Creating processes that enable and facilitate pragmatic and innovative \ourdomainshort{} designs.}
The diversity of already existing design processes (see also Section~\ref{sec:related}) can come as a surprise to the \ourdomainshort{} designer, and even to the best of system designers~\cite[Part I]{design:book/Brooks10}. Yet, the challenge of designing the \ourdomainshort{} design remains open. 

First, as we explain in Section~\ref{sec:related}, much exploration, combination, and innovation is still possible. The framework we propose in this work has been tested only by one research group, albeit large and long-lasting; new designs of (\ourdomainshort{}) design could prove vastly superior. 

Second, as the following challenges indicate, we have not yet understood the full extent of the problem raised by \ourdomainshort{} design. We envision new aspects will become relevant, leading to a co-evolving problem-solution.

\ \\
\challenge{ch:gooddesign}{\refpr{pr:design}}{Understand what is good design.}
Currently, the community relies largely on human experts to assess and curate designs. (In contrast, in hardware design, design space exploration has been largely automated.) 

{\it What is good design?}
We pose as an open challenge understanding (automatically) what is good design. This is not easy.

First, top venues use criteria such as ``degree of innovation'' and ``quality of the approach'', but their discrete formulation may ask reviewers to overfit their assessment to a quantitative estimation. Consequently, as exemplified in Figure~\ref{fig:violin-merit}, many scores cluster around the middle of the given range, leading to difficulties in separating the better designs from their near-equivalents. What alternative approach could be used?

Second, reviewers often also introduce in their assessment other criteria that have never been analyzed thoroughly. For example, simple designs are valued, which seems reasonable because simple designs foster system maintainability; but the evidence simplicity is the right trade-off between the quality of the approach and maintainability, or even a common understanding of what makes a system simple, are lacking. Other criteria, such as balance of the approach or another (semi-)aesthetic aspect (e.g., ``elegant design''), have also not been studied. This contrasts with the nature of real-world ecosystems, which are messy by nature, and which combine various designs created by different organizational cultures.

{\it How to assess good design?}
An already existing, albeit incomplete and largely subjective body of work facilitates starting our work on this challenge.

First, Altshuller discusses five large levels of design (``Levels of Creativity''~\cite[Part 1-2]{book:design:TRIZ}), evaluating them from the perspective of long-term, overall impact as: 
(1) trivial design, that is, using an existing design and minimally adapting it for local situations; 
(2) normal design, that is, selecting one of several designs, and adaptating the selected design after careful reasoning;
(3) novel design, that is, entailing significant adaptation of an existing design;
(4) fundamental design, that is, development of a new design or important feature, or the complete adaptation of an existing design (e.g., big data, serverless computing); 
and (5) outstanding design: a completely new ecosystem leading to significant scientific/technical advance (e.g., Maxwell’s electricity laws and first use in practice, the Internet, the cloud).
Alternatives to this set of levels exist, but roughly follow the same structure, e.g., the rating systems for top conferences roughly consider levels (1) through (4) in Altshuller's taxonomy.

Second, Altshuller also discusses four levels of design, from the perspective of performance against alternatives, vs. random design, na\"{i}ve design, current practice, and ideal or optimal alternatives. Other frameworks exist, but these levels are typically considered by reviewers when assessing the technical quality of the experimental setup.

Third, especially the academic community has proposed some quantitative measures for quantifying the creativity and effectiveness of designs in fields with rather narrow design spaces~\cite{design:metrics:ShahSV03,design:metrics:SarkarC11}.
How these or related metrics could be put in practice for \ourdomainshort{} design, and what metrics remain to be invented, is unknown.

\challenge{ch:exploration}{\refpr{pr:design}}{Simulation-based approaches and experimentation for design space exploration. Calibration and reproducibility are key.}
In maze-solving, it is known that finding an exit is much harder when the alternatives are numerous than for a straight path. Yet, in our field, the complexity and the number of alternatives considered and eliminated before the design has emerged, or more broadly the characteristics of the design space, are rarely discussed in our articles or by their reviewers. How to characterize the broad and diverse design spaces available in \ourdomainshort{} design?

\subsection{Systems Challenges}
\label{sec:challenges:system}
\label{sec:ch:system}

\challenge{ch:ecosystems}{\refpr{pr:ecosystems}}{Design for \ourdomainshort{}, not for individual systems.}
We see as the grand challenge of \ourdomainshort{} design to understand how the resulting design will fit in an entire ecosystems. Typical questions include: How to enable and how to future-proof the design of systems that need to interoperate, especially dynamically, at runtime? For example, how to enable cross-cloud operation, service delegation and federated composition, and geo-distributed data use? Is this even achievable with high likelihood of success, when ecosystems combine organically designs from different organizations and business units, and thus suffer the consequences of Melvin Conway's empirical law~\cite{design:laws:Conway68} that designs ``by committee'' are likely to fail?

Current approaches already reveal patterns in the core topics pursued by the community. 
These include~\cite{DBLP:conf/icdcs/IosupUVAEHTBT18}: 
(i) adaptation and self-awareness in ecosystems, 
(ii) ecosystem navigation: find and solve common problems of comparison, selection, composition, replacement, adaptation, and operation; 
(iii) discovering the new world: creating designs responding to new modes of use;
(iv) the challenge to support non-functional requirements (see \refpr{pr:nfr}{});
(v) the ecosystem-scheduling challenge: design scheduling approaches to be flexible enough to represent \ourdomainshort{} needs, diversity, and heterogeneity, and solve both the provisioning and the allocation problems.

Addressing this challenge could also start from understanding the workload and relative importance of individual components in current ecosystems. This could give quantitative evidence that some components are naturally more important than others, and thus focus the community efforts. One of the likely steps in this sense is to observe pragmatically which part of the current ecosystem is taking much engineering time, and re-design that part into ``$\star$-as-a-Service''.

\ \\
\challenge{ch:catalog}{\refpr{pr:ecosystems}}{Establish a catalog of components for \ourdomainshort{} design.}
Such a catalog would consist of design principles, known architectural and operational patterns, etc. Useful catalogs are a known approach for settled fields~\cite{design:book/BeitzPFG07,book:design:BellMM78}, but how to build a useful catalog for \ourdomainshort{} designs?

\subsection{Peopleware Challenges}
\label{sec:challenges:peopleware}
\label{sec:ch:peopleware}

\challenge{ch:curriculum}{\refpr{pr:educating}}{Create a teachable common body of knowledge for \ourdomainshort{} designs, focusing on pragmatism, innovation, and ethics. Design effective teaching practices for this curriculum.}
{\it How to teach design for \ourdomainshort{}?} 
From traditional courses on distributed systems, we know that students face 
many daunting technologies, such as clouds, clusters, and peer-to-peer, and 
should learn about many conceptual advances in very diverse topics, such as the functional aspects of consistency, synchronization, commit and agreement, etc., and the non-functional aspects of performance, dependability, security, etc.
More recently, we have started to understand that students trained with the traditional approach lack an understanding of how the concepts have evolved~\cite{DBLP:journals/cacm/Gal-EzerH98,DBLP:journals/cacm/Tekir12,book:philosophy:Tedre15}. 
We propose that the body of knowledge for \ourdomainshort{} design should include the history and evolution of designs, key debates in design (e.g., end-to-end vs. local properties), evolving architectures and patterns, etc.

To avoid another ``Image Crisis'', we also need to include in our curricula elements of ethics~\cite{DBLP:journals/cacm/Vardi19}. However, one could argue that ethics courses have been added already into current curricula. We see these courses more as shoved than added, and posit that ethics courses for this field should be presented from the perspective of experts in the \ourdomainshort{} field, and not as abstract concepts presented from the remote lens of general philosophy.

{\it The Distributed Systems Memex:} 
In the 1940s, Vannevar Bush defined the concept of the personal {\it memex} as a person's device for storing and accessing all information and communication involving that person~\cite{memex45}. 
Bush identifies many benefits for archiving large amounts of personal data into the memex, including learning about and eradicating diseases, enabling more creative and thought-related time by eliminating tasks
that can be automated, etc. 
(Bush does not spend much time on the drawbacks, which eventually led to the privacy-related regulations, e.g., GDPR.)
Similarly, we have posited~\cite{Iosup:memex12} that archiving large amounts of operational traces collected from the distributed systems that currently underpin our society can be highly beneficial for \ourdomainshort{} design. 
Even the design of such a the Distributed Systems Memex is non-trivial, and may teach us much about the key operational principles of distributed systems and ecosystems. 
What should the Distributed Systems Memex include? What data? Which types of distributed systems? How can such a Memex be designed? What instruments could it use and how could it be implemented overall? 

We see now an additional aspect of the Distributed Systems Memex: the preservation of original designs and of their origins. We are losing valuable heritage by not preserving the artifacts of design, the decisions that lead to them, and the thoughts and discussions that led to these designs; capturing these later may not be possible much later, as the generation that produced them in the 1980s, 1990s, and 2000s will start to retire.

\challenge{ch:engagement}{\refpr{pr:educating}}{Create communities and environments for people to engage with the design and operation of ecosystems, to demonstrate and explain operations.}
As demonstrated by the success of Facebook over its competitors, communities must be supported by proper interactive tools. Besides the tools for social networking, we envision for this challenge:
(1) support for showing and explaining the operation of ecosystems, and of their constituent systems, to all stakeholders, continuously;
(2) tools to demonstrate and explain the impact of various design principles used in the  design of distributed ecosystems and of their systems, to a diverse community;
(3) tools to explore the impact of various design trade-offs used in \ourdomainshort{} design, aiming to support a diverse community;
etc.
\ \\

\subsection{Methodological Challenges}
\label{sec:challenges:methodology}
\label{sec:ch:methodology}

\challenge{ch:documenting}{\refpr{pr:science}}{($\star$) Design a formalism for documenting designs.}
{\it How to trace the evolution of designs?}
An open process for design requires more than its final results and artifacts to be made public. Whatever the reason, many design-decisions happen behind closed doors and are never revealed. 
Designs also incorporate intangibles, such as the experience leading the designer to take specific design decisions. These factors make the provenance of design choices difficult to track. Compounding the problem, documenting the provenance requires formalisms and description languages that should not hamper the creative process, and in particular should not punish extensive and creative designs.

{\it How to validate the views on the evolution of designs?}
In general, we have very limited examples documenting the evolution of designs. The most comprehensive in this sense are the early and heroic efforts by former employees of DEC and IBM, who tried to capture how the field of hardware computer systems has evolved~\cite{book:design:BellMM78,book:design:BlaauwB97}. The two resulting books capture tens of designs, each in its own way---Bell et al. through a formalism, Blaauw and Brooks through a historical evolutionary graph. Our own recent work on serverless computing~\cite{DBLP:journals/internet/EykTTVUI18} uses the latter approach to capture the technology leading to serverless computing over multiple decades; however, the community has still not started to debate these structured histories, and the possibly alternative views captured by them.

\ \\
\challenge{ch:ethno}{\refpr{pr:evolution}}{Understand \ourdomainshort{} design in practice. How and when do \ourdomainshort{} practitioners design what they design?}
We know normal abduction is commonly used in engineering~\cite[Ch.5]{design:book/Arthur09}~\cite[Part I]{design:book/Brooks10}, especially coupled with complex implementation and realization processes~\cite{book:design:BlaauwB97}.
However, the extent and approach of using design abduction in practice are not currently known. 
The ``When?'' is also important; for example, design used to be a static process done at the start of projects, but in some dynamic organizations it is now part of the weekly sprints and helps the DevOps teams respond quickly. In consultancy, design may encounter strict time constraints, and also need to address unusual requirements, as follows.

{\it Design for datacenter operators:} 
Many cloud operators use design to ecosystems that are highly reliable but still flexible. Whereas requirements solicitation traditionally would take place only at the beginning of a project, such processes are now an integral part of cyclic design. New feature requests from customers and findings from operations drive cloud operators to revisit earlier designs and change them to accommodate new technology, reliability improvements, new features, etc.
Design in cloud operations helps find solutions for NFRs, to solve challenges of reliability, maintainability, and security.
Design used to be a static process at the start of projects, but it is now likely part of weekly sprints, and thus helps the DevOps teams respond  quickly to a fast changing IT landscape.

{\it Design for consultancy} is complex, as for many projects there is a rigid time constraint coupled with unusual requirements. Every design needs to be custom-tailored to the needs of each customer. The design process must cover the full spectrum of IT transformation: from choosing the right hardware to delivering working software. Unlike other businesses, a consultancy also has to design the organizational change required for a client to maintain and operate the finished product (peopleware: types of human resources, required skills, knowledge transfer, etc.). Attracting customers involves working at the cutting edge of technology; however, many customers have legacy systems that need to integrate with the new technology (see also Section~\ref{sec:motivation:mcs}). Likely, the design process at a consultancy begins with a well-formulated initial design that covers the requirements from all stakeholders, with short iterations after discussions with the client. The focus is on automation and on customer self-service, to best support requirements such as availability, reliability, maintainability, usability, and security. 
To support integration with various legacy systems and with new ecosystems, solutions must be flexible and API-driven. Infrastructure designs cover the entire spectrum from private cloud to hybrid and public cloud solutions.

\ \\
\challenge{ch:orgsim}{\refpr{pr:evolution}}{Organizational similarity in \ourdomainshort{} design.}
Given how \ourdomainshort{} designs are likely to span multiple designers and thus also organizational cultures, it is surprising to see how little detail appears in the published articles about the environment in which the designs were produced. It would be interesting to look for evidence of {\it organizational similarity} across the designs originating in largely similar organizations. Conversely, it could be valuable to consider the designs originating under different organizational cultures.

%% file: content/06-design-exp.tex
\ \\
\newpage

\begin{table}[!t]\centering
\begin{adjustbox}{max width=\columnwidth}
    \ra{1.1}
    \begin{tabular}{@{}lll@{}} \toprule
        Section & Experiment & Key aspects \\ 
        \midrule
        $\S$\ref{sec:exp:p2p} & P2P & Protocol/Sys. design \\
        $\S$\ref{sec:exp:mmog} & MMOG & Ecosystem, NFRs\\
        $\S$\ref{sec:exp:dcmgmt} & DC management &  RM\&S, ref.archi. \\
        $\S$\ref{sec:exp:serverless} & Serverless, FaaS & Design in new ecosystem \\
        \midrule
        $\S$\ref{sec:exp:graphalytics} & Graphalytics & Ecosystem design, Laws \\
        $\S$\ref{sec:exp:portfolio} & Portfolio scheduling & System design \\
        $\S$\ref{sec:exp:autoscaling} & Autoscaling & Experiment design \\
        \bottomrule
        \phantom{abc}\\
    \end{tabular}
\end{adjustbox}
\vspace*{-0.25cm}
\caption{Experiments with the \AtLarge{} design framework.}
\label{tab:exp}
\label{tab:exp:overview}
\vspace*{-0.5cm}
\end{table}

\section{Experiments with the Design Framework}
\label{sec:use}\label{sec:exp}\label{sec:designexperiments}\label{sec:designexp}

We have used the \AtLarge{} design framework as our main approach to design for the past decade. Effectively, we have designed its various processes, conducted experiments with them, and refined them as we uncovered the many problems a research group faces in creating pragmatic and innovative \ourdomainshort{} designs. (Our designs are also ethical, both in our view, and as assessed by our reviewers, institutions, and funding agencies.)

We summarize in Table~\ref{tab:exp:overview} our use of the \AtLarge{} design framework for over a decade, for a broad range of \ourdomainshort{} designs:
\begin{enumerate}
    \item Co-evolving understanding, and protocol and system design for Peer-to-Peer systems~(Section~\ref{tab:exp:p2p});

    \item Design for online gaming ecosystems~(Section~\ref{tab:exp:mmog}), as an example of designing in rapidly changing ecosystems operating under strict NFRs;

    \item Design for datacenter ecosystems~(Section~\ref{sec:exp:dcmgmt}), as an example of evolving understanding of how the field's reference architecture is emerging;

    \item Design for serverless ecosystems~(Section~\ref{sec:exp:serverless}), as an example of how the \AtLarge{} design process fosters collaboration between diverse teams;

    \item The design of the Graphalytics ecosystem~(Section~\ref{sec:exp:graphalytics}), as an example of DevOps support;
    
    \item Design for portfolio scheduling~(Section~\ref{sec:exp:portfolio}), as an example of co-evolving a detailed system-level design;

    \item The design of experiments in autoscaling~(Section~\ref{sec:exp:autoscaling}), as an example of designing both real-world and simulation-based experiments.
\end{enumerate}

Overall, we conclude the \AtLarge{} design framework passes the following criteria for success: 
\begin{enumerate}
    \item It allow us to co-evolve problems and their solutions, even for problems with very successful solutions, or for very challenging problems with no or few solutions;
    
    \item It help us identify ``hot'' problems, and make scientific discoveries with impact on the community;
    
    \item It enables us to create pragmatic and innovative designs, as assessed by our own team and by the expert reviewers;
    
    \item It keeps our design activity fit to receive competitive funding from academic and industrial funding organizations, and interesting and motivating to attract a diverse group of young researchers eager to challenge the new problems;
    
    \item It results in publications accepted by high-quality venues, which we see as proxies of high-quality designs and results, and foster other useful results ancillary to good design practices (e.g., publishing high-quality software and data artifacts).
\end{enumerate}

We now address each of the seven types of \ourdomainshort{} design activities from Table~\ref{tab:exp}, in turn.


\input{content/exp-p2p.tex}

\input{content/exp-mmog.tex}

\input{content/exp-dcmgmt.tex}

\input{content/exp-serverless.tex}

\input{content/exp-graphalytics.tex}

\input{content/exp-portfolio.tex}

\input{content/exp-autoscaling.tex}

%% file: content/exp-p2p.tex
\begin{table}[!t]
  \caption{Co-evolving problem-solutions in our P2P work.}
  \label{tab:exp:p2p}
    \begin{adjustbox}{max width=\columnwidth}
  \begin{tabular}{lllll}
    \toprule
    Study & Feature & Depth & Breadth & Instruments\\
    \midrule
    \multicolumn{5}{c}{{\it Longitudinal studies}}\\
    \midrule
    \cite{conf/asci/IosupGPE05} ('05) & Aliased media & Low & Wide & Analytics \\
    \cite{DBLP:conf/ccgrid/IosupGPE06} ('06) & Ecosystem-Internet & Deep & Wide & MultiProbe \\
    \cite{DBLP:conf/hpdc/WojciechowskiCPI10} ('10) & Global ecosystem & Deep & Wide & BTWorld \\
    \midrule
    \multicolumn{5}{c}{{\it Other studies and instruments}}\\
    \midrule
    \cite{TR:P2PTA:Zhang2010} ('10) & P2P Trace Archive & Low & Wide & Analytics \\
    \cite{DBLP:conf/europar/ZhangIPES10} ('10) & Bias & Deep & Wide & Analytics\\
    \cite{DBLP:conf/p2p/ZhangIPE11} ('11) & Flashcrowds & Deep & Wide & Analytics\\
    \cite{DBLP:conf/bigdataconf/HegemanGCHEI13} ('13) & Global ecosystem & Deep & Wide & BTWorld \\
    \cite{DBLP:conf/ccgrid/GhitCHHEI14} ('14) & Vicissitude & Deep & Narrow & BTWorld \\
    \midrule
    \multicolumn{5}{c}{{\it New P2P features and systems}}\\
    \midrule
    \cite{DBLP:conf/p2p/GarbackiIES06} ('06) & Collaborative & Deep & Narrow & 2fast \\
    \cite{DBLP:journals/concurrency/PouwelseGWBYIERSS08} ('07) & Social & Deep & Wide & Tribler \\
  \bottomrule
\end{tabular}
\end{adjustbox}
\end{table}

\subsection{The Design of P2P Systems} \label{sec:exp:toofast} \label{sec:exp:p2p}

We present in this section an overview of our design work in Peer-to-Peer~(P2P) computing. 
The approach of co-evolving problem-solution led us to new insights into the operation of P2P systems and to innovative new features and systems, as summarized by Table~\ref{tab:exp:p2p}. It has also allowed us to participate in an exciting moment of high-paced evolution in understanding and designing distributed systems.

P2P computing is a paradigm under which participating entities in a distributed system (the {\it peers}) can use direct, two-way (peer-to-peer) communication to perform and/or receive some service. The ability to communicate directly allows peers with the desire to provide and/or use a specific service ({\it similar interest}) to group ({\it swarm}). Because peers can both perform and receive service, peer-to-peer systems promise to use all the available resources, be available as long as even one peer survives, and scale up with the resources volunteered by peers even during high-intensity periods ({\it flashcrowds}). Thus, P2P systems promise very desirable properties, such as high performance, availability, and efficiency, and also cost-effectiveness (as a {\it near-zero-cost technology}).

We trace the origins of our work in P2P to our collaboration with a team, Pouwelse et al., finalizing their 8-month investigation of the BitTorrent file-sharing network~\cite{DBLP:conf/iptps/PouwelseGES05}. At the time, there existed only a handful of measurement studies and only a few theoretical treatments of BitTorrent~(BT); each had significant shortcomings. The study led by Pouwelse et al. was indeed seminal, providing longitudinal data about the global SuprNove BT-ecosystem, and uncovering for example a real-world flashcrowd and debunking theoretical assumptions such as Poisson arrivals. 

The Pouwelse et al. study started a community frenzy, a race of a type common in natural sciences, to complete the first or the largest study of previously unknown phenomena.
We joined this BT-related frenzy around October 2004, which was the time the Pouwelse et al. were wrapping up their study. Our team was afforded generous, unfettered access to the original data collected by Pouwelse et al., resulting in a series of deepening studies, complemented by new measurements and by broader engagement of the community to create a Peer-to-Peer Trace Archive~\cite{TR:P2PTA:Zhang2010}. 

Our studies have uncovered several ecosystem-level phenomena, such as:
(1) the 2005 analytics study~\cite{conf/asci/IosupGPE05}, resulting in the discovery of the presence of {\it aliased media}, which is the presence of very similar media content in a variety of formats in the global BT-ecosystem, and the first characterization of aliased media operation;
(2) the 2005 longitudinal study of the global PirateBay BT-ecosystem correlated with Internet-level measurements~\cite{DBLP:conf/ccgrid/IosupGPE06}, which has uncovered that 
the bandwidth capacity of BT-users has shifted to a large imbalance between upload and download (due to widespread adoption of ADSL technologies) and, in the race, has remained the largest and most comprehensive BT study until 2009;
(3) the 2010 deep study of the global BT-ecosystem~\cite{DBLP:conf/hpdc/WojciechowskiCPI10}, collected nearly 1 billion samples across hundreds of trackers and over 10,000,000 BT-swarms, and revealed the existence of {\it giant swarms} of hundreds of thousands of concurrent users, of {\it spam trackers} inserted by unidentified entities to presumably mislead and track BT-users, and in general of a robust global BT-ecosystem;
(4) the 2011 study of BT-flashcrowds~\cite{DBLP:conf/p2p/ZhangIPE11}, for which we developed a method to identify flashcrowds, the first comprehensive model of BT-flashcrowds, and showed evidence of important negative phenomena that occur only during flashcrowds. 
These studies also led us to (5) meta-analysis~\cite{DBLP:conf/europar/ZhangIPES10}, that is, to study the systematic bias introduced by the measurement instruments, and to catalog and characterize various sources of bias.
Our BT-studies also led us to create (6) the Peer-to-Peer Trace Archive for sharing data publicly, as FOAD (see Section~\ref{sec:core:solveproblem}).
Last, and as a warning to young researchers, (7) these studies revealed to us the importance of good design for reporting: because our reporting skills have not yet refined, and because we lacked a structured process to compensate, our work prior to 2010 got rejected repeatedly when submitted to major systems and networking conferences. (Writing seems to be an important reason, because our discoveries are on-par with the phenomena described in the accepted articles of the period, and are based on at least as deep and as comprehensive factual evidence.)

Each of these studies required the development of new systems for measurement and analysis, including MultiProbe~\cite{DBLP:conf/ccgrid/IosupGPE06} and BTWorld~\cite{DBLP:conf/hpdc/WojciechowskiCPI10}, which are both global-scale monitors for BT-ecosystems, the former also focusing on collecting Internet-tracing data (not possible anymore under GDPR laws), the latter focusing on efficient collection of aggregate-data.
In 2014, while trying to analyze the full BTWorld-dataset, we developed a novel big data analytics pipeline~\cite{DBLP:conf/bigdataconf/HegemanGCHEI13}; the process allowed us to discover the phenomenon of {\it vicissitude}~\cite{DBLP:conf/ccgrid/GhitCHHEI14} (see Section~\ref{sec:motivation:mcs}).

The phenomenon of upload-download bandwidth asymmetry in BT-ecosystems led us to design 2fast~\cite{DBLP:conf/p2p/GarbackiIES06}, a BT-compatible protocol for collaborative downloads where the incentive to share does not need immediate repay and thus can lead to efficient use of asymmetric bandwidth. In particular, we showed 2fast serves not only a social function, but also can improve significantly the performance of BT-based file-sharing. Between 2005 and 2010, 2fast was one of the three main pillars of the first socially aware P2P system, Tribler~\cite{DBLP:journals/concurrency/PouwelseGWBYIERSS08}, and is thus partially responsible for the nearly 500,000 downloads recorded by Tribler in that period.

%% file: content/exp-mmog.tex
\begin{table}[!t]
  \caption{Co-evolving problem-solutions in our MMOG work.}
  \label{tab:exp:mmog}
    \begin{adjustbox}{max width=\columnwidth}
  \begin{tabular}{lllll}
    \toprule
    Study & Feature & Depth & Breadth & Instruments\\
    \midrule
    \multicolumn{5}{c}{{\it Longitudinal studies}}\\
    \midrule
    \cite{DBLP:conf/sc/NaeIPPEF08} ('07) & Dynamics & Low & Wide & Runescape \\
    \cite{DBLP:conf/have/GuoSVI12} ('12) & Dynamics & Low & Wide & MOBA \\
    \cite{DBLP:conf/wosp/OlteanuIT13} ('13) & Dynamics & Low & Wide & Social \\
    \cite{DBLP:journals/internet/IosupBSJK14} ('13) & Soc.nets. & Deep & Wide & Social \\
    \cite{DBLP:journals/tomccap/JiaSEI16} ('16) & Soc.nets. & Deep & Wide & Meta-gaming \\
    \midrule
    \multicolumn{5}{c}{{\it Other studies}}\\
    \midrule
    \cite{DBLP:conf/netgames/ShenVI11} ('11) & Scaling & Deep & Narrow & RTSenv \\
    \cite{DBLP:conf/netgames/MartensSIK15} ('15) & Toxicity & Deep & Wide & Social \\
    \midrule
    \multicolumn{5}{c}{{\it New MMOG features and systems}}\\
    \midrule
    \cite{DBLP:conf/sc/NaeIPPEF08} ('07) & V-World & Deep & Wide & MMOG \\
    \cite{DBLP:conf/europar/Iosup09} ('09) & PGCG & Deep & Wide & POGGI \\
    \cite{DBLP:conf/netgames/IosupLT10} ('10) & Analytics & Deep & Wide & CAMEO, cloud \\
    \cite{DBLP:conf/wosp/NaePIF11} ('11) & V-World & Low & Wide & SLAs, Business \\
    \cite{DBLP:journals/tomccap/ShenHIE15} ('15) & V-World & Deep & Narrow & Scalability \\
    \cite{DBLP:journals/concurrency/JiangVPI18} ('18) & V-World & Deep & Narrow & Mirror \\
    \midrule
    \multicolumn{5}{c}{{\it Software Instruments and Data Archives}}\\
    \midrule
    \cite{DBLP:conf/netgames/GuoI12} ('12) & Archive & Low & Wide & GTA \\
    \cite{DBLP:conf/wosp/SarDI19} ('19) & Benchmark & Deep & Narrow & Yardstick \\
  \bottomrule
\end{tabular}
\end{adjustbox}
\end{table}

\subsection{Design for MMOG Ecosystems} \label{sec:exp:mmog}

We present in this section an overview of our design work in Massive Multiplayer Online Games~(MMOG), which is a popular and lucrative application domain of \ourdomainshort{}. 
Similarly to our design work for P2P systems (see Section~\ref{sec:exp:p2p}), for MMOG we used the approach of co-evolving problem-solution for over a decade, understanding and improving the operation of MMOG ecosystems, as summarized in Table~\ref{tab:exp:mmog}. 
But MMOGs are not yet another type of distributed system, for two main, systems-related reasons: they raise uniquely challenging NFRs, and they have evolved significantly over the past decade. 
Thus, this section gives evidence that the \AtLarge{} design framework can deliver good designs in a challenging and rapidly evolving field.

MMOGs operate as very large\footnote{Around 2010, the popular MMOG World of Warcraft operated on a global distributed ecosystem of over 10 datacenters, with a total scale rivaling that of the computing grid supporting the Large Hedron Collider, one of the largest scientific experiments in the world.} and diverse ecosystems~\cite[$\S$6.3]{mcs18tr}, raising some of the strictest NFRs in distributed systems. They require continuous consistency, high-frequency updates, low performance variability, etc., while supporting scales of possibly millions of concurrent users connected to each other over the (performance-varying) Internet, and not losing the ability of a game operator to keep oversight of the game.
(This is unlike many P2P systems, where completing the service correctly seems much more important than meeting NFRs.) 

Our MMOG work coincided with a moment of great interest and diversification in gaming, experienced by both the systems community and the society at large. 
We know now that the
typical MMOG ecosystem combines four broad functions~\cite[$\S$6.3]{mcs18tr}: 
(1) the operation of the virtual world (V-World), which is the focus of classical systems work in online gaming but also raises numerous new challenges~\cite{DBLP:conf/icmcs/IosupSGHDP14}; 
(2) gaming analytics, which combines the collection and analysis of big gaming data, and the creation of actionable systems- and business-level decisions; 
(3) procedural game-content generation (PGCG), which combines computational and game-design challenges;
and (4) meta-gaming, which raises the challenge of operating a social network for gamers to share experiences, screenshots, and videos, and to discuss game tournaments and other issues.
However, at the start of our work in MMOG, 
the overall 
Function~(1) meant largely MMO Role-Playing Games (MMORPGs), and only later did large-scale First-Person Shooter (FPS) and Real-Time Strategy (RTS) games become MMOs, and did multiplayer online battle arena (MOBA) and online social (OS) games appear;
Function~(2) was in much use inside the major game operators, but there was relatively little use of big data technology and there were few other organizations doing gaming analytics;
Function~(3) was uncommon;
and Function~(4) was barely known, as the social networking market was still emerging and very fragmented.

Because of the vast design space, our design work for MMOGs can only be characterized as exploratory. 
We started with gaining a deep understanding of how these applications operate in practice, uncovering the short- and long-term dynamics of popular MMORPGs~\cite{DBLP:conf/sc/NaeIPPEF08}. We did this by tracing, from around 2005 to 2008, the operations of multiple MMORPGs, and in particular of one that became one of the most popular MMORPGs, Runescape.
This work led us to design techniques for resource management and scheduling for cloud- and  datacenter-based MMOG operations~\cite{DBLP:conf/sc/NaeIPPEF08,DBLP:journals/tpds/NaeIP11}; our efforts were followed by independent designs going in the same direction~\cite{DBLP:conf/IEEEcloud/LeeC10}.
We also combined in our design both technology and business considerations~\cite{DBLP:conf/wosp/NaePIF11}\footnote{Combining technology and non-technology considerations, such as business and creative, makes the resulting work better fit to solve problems, but cases considerable problems to reviewers. For example, our work combining the technology and business of MMOG was rejected repeatedly until finding a community willing to consider such diverse aspects~\cite{DBLP:conf/wosp/NaePIF11}.}.
Having understood from our work that MMORPGs based on clouds could scale elastically, almost ``by credit-card'', we turned our attention to how MMOGs operated outside of their main Function~(1), with innovative designs for PGCG, for which we invented the first distributed and parallel system to generate fresh and diverse content at scale~\cite{DBLP:conf/europar/Iosup09}\footnote{POGGI won a distinguished paper award from Euro-Par, in 2009.} and for cloud-based analytics, for which we combined NoSQL and cloud technology to design one of the first systems for gaming analytics at scale~\cite{DBLP:conf/netgames/IosupLT10}\footnote{Leading MMOG companies, such as Blizzard, started to discuss similar approaches publicly around 2016, nearly 7 years after our CAMEO publication.}.

Our early work with MMORPGs made us ask the important question of whether we could scale to MMORPG-like scales the existing RTS games, which have significantly more challenging NFRs (e.g., lower latency and stricter consistency). We started with trying to gain a deeper understanding of why existing RTS games fail to scale, and designed the first benchmark for this purpose, RTSenv~\cite{DBLP:conf/netgames/ShenVI11}. By applying RTSenv to one of the few RTS games avaiable as FOSS, we discovered a new form of scalability, unique to MMOGs, that combines systems and game-design concepts~\cite{DBLP:conf/netgames/ShenVI11}. 
The consequence of this discovery is that simply scaling RTS games by scaling their technology but without taking into account the interactive details\footnote{Although the principles of computer-human interaction are well-understood, for the practice of MMOG design they give guidelines rather than quantitative, actionable information. In this, they play a similar role to how the laws of physics act on hardware design.} of how they are used (e.g., where units are located, how many actionable items appear on the same screen). 
This made us change the focus of our work on scalability, from traditional problems, to scaling based on how (MMORTS) games are used.
We understood we had to address not only the limitations revealed in the lab by RTSenv, but also new problems derived from how players actually interact with their games. 
To this end, we found out that the gaming community was already collecting data about real-world use, but for training purposes---professional, semi-professional, and amateur players learned from the best-performers by replaying their game sessions---, in an early example of the complexity of meta-gaming operations. 
By analyzing the game replays, we found out that RTS games, unlike MMORPGs, (i) have multiple points of interest, (ii) require careful management of up to tens of entities in some of the points, and (iii) require more casual management of up to hundreds of entities in the others; this resulted in the design of the Area of Simulation MMOG-technique and system~\cite{DBLP:journals/tomccap/ShenHIE15}, and, for cloud-based operation, of the Mirror system that can offload computation~\cite{DBLP:journals/concurrency/JiangVPI18}.

The discovery of MMORPG-related phenomena made us curious about exploring the MMOG universe more broadly, about uncovering the properties of more, if not all major types of MMOGs. (The parallel we can draw from the more conventional work on parallel and distributed systems is with the U.C.-Berkeley ``views of'' series~\cite{DBLP:journals/cacm/AsanovicBDKKKMPSWWY09,DBLP:journals/cacm/ArmbrustFGJKKLPRSZ10}.)
Over the next few years, we uncovered 
the short- and long-term dynamics of 
MOBA~\cite{DBLP:conf/have/GuoSVI12} and 
OS~\cite{DBLP:conf/wosp/OlteanuIT13} games, 
and new, deeper phenomena occurring in emerging and more established game genres.

Among the deeper phenomena we have discovered are 
the implicit social-network forming in various kinds of game genres~\cite{DBLP:journals/internet/IosupBSJK14} and in meta-gaming~\cite{DBLP:journals/tomccap/JiaSEI16}. 
Their importance remains underestimated, and our work in design joins a small but emerging body of work focusing 
both on positive applications such as matchmaking~\cite{DBLP:journals/internet/IosupBSJK14,DBLP:journals/tkdd/JiaSBIKE15} and best-practice sharing~\cite{DBLP:journals/tomccap/JiaSEI16}, and on
preventing negative situations such as online bullying and toxicity~\cite{DBLP:conf/netgames/MartensSIK15}.

Because online gaming technology raises challenging NFRs, in our experience prototypes in this domain take much longer to develop than, for example, P2P designs. For example, implementing the Area of Simulation system~\cite{DBLP:journals/tomccap/ShenHIE15} took over 6 months. 
To keep such software development projects under control, it was vital to turn to using good software design processes. 
We learned an important lesson.
Prototypes are essential in testing NFRs, because current software design practices do not lead to sufficient guarantees of, e.g., low latency. Unfortunately, in our experience, reviewers of designs with strict NFRs do not have a good understanding of the challenges posed by prototyping, and tend to dismiss such prototypes as not important for the act of design especially for emerging application domains. We challenge this culture of reviewing designs through this work.

We learned another lesson through our design work. 
One of the key contributions a team can make to the field is, in our view, sharing workload and operational traces in a FAIR and/or FOAD archive, such as the Game Trace Archive~\cite{DBLP:conf/netgames/GuoI12}.
The design of such an archive is always pragmatic, but in terms of innovation it can only achieve as much as required by the type of data to be shared in the archive. 
We challenge that reviewers should understand better the standards of innovation in creating data archives, and judge innovation here also by the novelty of the actual data shared through the archives.

%% file: content/exp-dcmgmt.tex
\begin{figure}[!t]
    \centering
    \includegraphics[width=1.0\columnwidth]{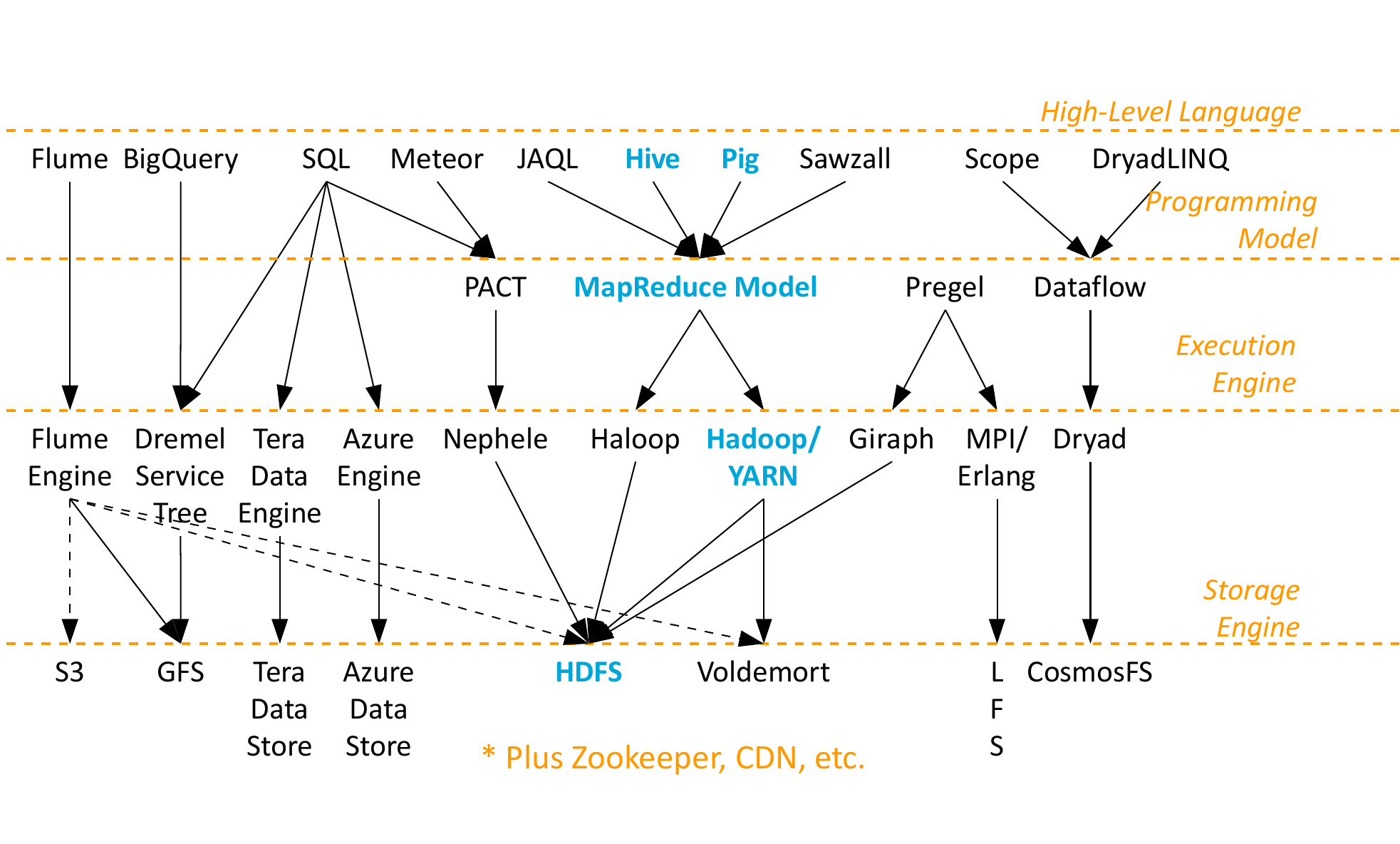} 
    \includegraphics[width=1.0\columnwidth]{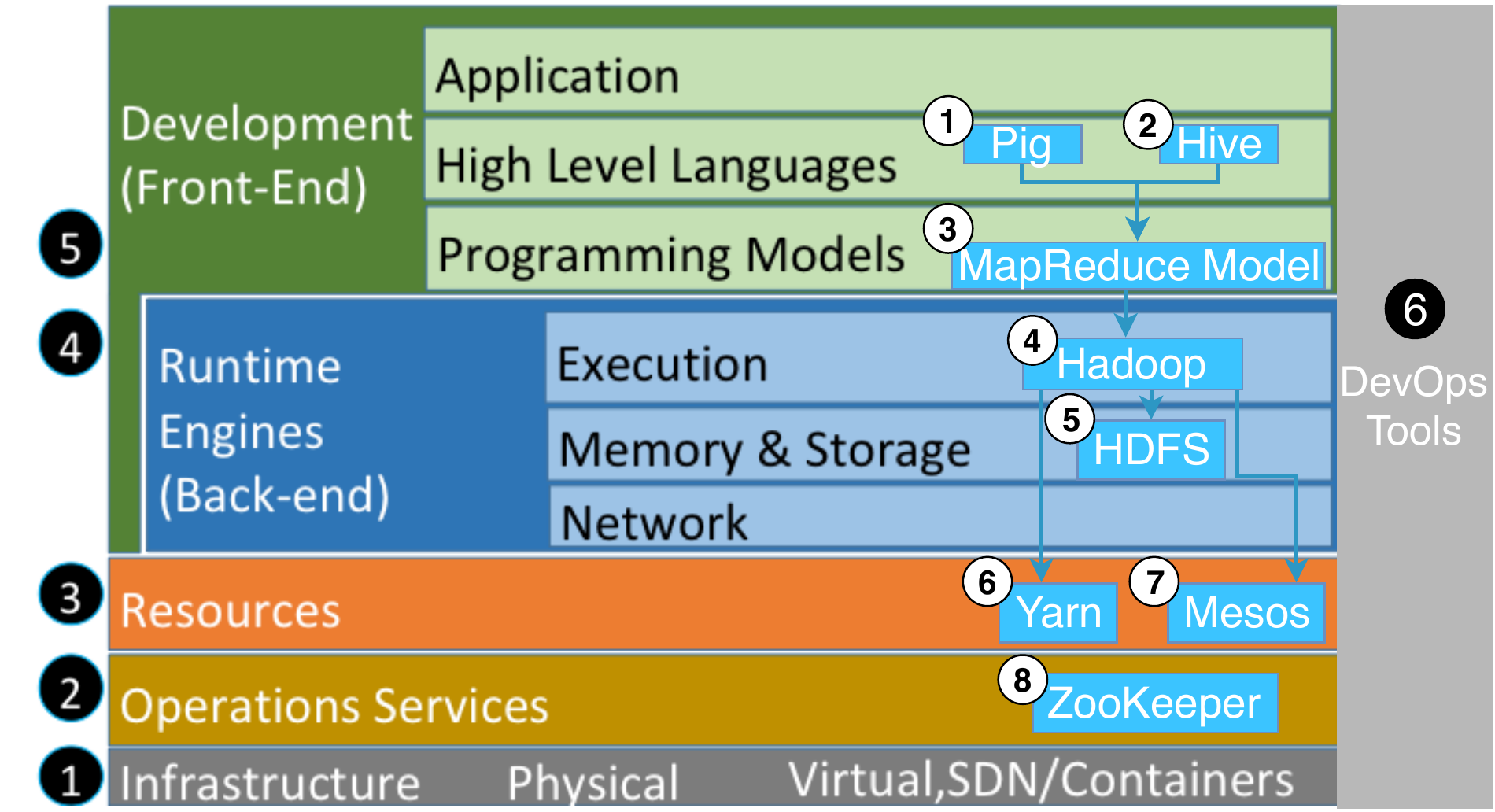}
    \caption{An evolving reference architecture for the datacenter ecosystem: ({\it top}) Our understanding of the big data ecosystem, 2011--2016. Reproduced and adapted from our previous work~\cite{DBLP:conf/ccgrid/GhitCHHEI14}.
    ({\it bottom}) Our understanding of the entire datacenter ecosystem, 2016--ongoing. Includes a sample mapping of a MapReduce-based big data ecosystem.}
     \label{fig:refarchidc}
    \vspmed
\end{figure}  

\subsection{Datacenters: Designing the Digital Factory}
\label{sec:exp:dcmgmt}

We present in this section the evolution of a reference architecture for the ecosystems operating in the datacenter. 
A reference architecture facilitates the design of systems, stacks, and platforms, allowing the designer to start from an overview of how the entire ecosystem works. 
Figure~\ref{fig:refarchidc} depicts both our initial design focusing on the big data ecosystem, and the revised and extended design for the entire datacenter ecosystem.

Datacenters hold a crucial place at the heart of the Digital Economy, producing efficient, dependable services. They allow clients to run diverse workloads, including data processing pipelines, 
scientific simulations, and online gaming, all with the promise to achieve efficiency and near-optimal resource utilization. In addition to this challenge, we can identify another one that addresses the diversity of infrastructures. Datacenters appear in different scales and designs, from multi-cluster deployments like Amazon EC2 and Microsoft Azure to cloud-edge~\cite{DBLP:journals/computer/Satyanarayanan17} micro-datacenters used for video trans-coding and streaming~\cite{DBLP:journals/computer/Ananthanarayanan17}. This raises numerous scientific, design, and engineering challenges~\cite[$\S$6.1]{DBLP:conf/icdcs/IosupUVAEHTBT18}.

Our initial design of a reference architecture for datacenter-based ecosystems started in 2011, with a drawing of big data ecosystems created jointly by the community in a high-profile Dagstuhl Seminar. For nearly 5 years, we have refined that drawing and added to it our own understanding of the topic.  Figure~\ref{fig:refarchidc}~(top) depicts the resulting reference architecture for big data. The four layers, {\it High-Level Language}, {\it Programming Model}, {\it Execution Engine}, and {\it Storage Engine}, are conceptual, but applications that run in this ecosystem typically use components across the full stack of layers (and more, as indicated by the $\star$ in the figure). The highlighted components cover the minimum set of layers necessary for execution for the MapReduce ecosystem; the presence of several high-level languages indicates that the ecosystem has diverse users, with minimal expertise and ability in managing the ecosystem beyond the high-level language they know. 
This reference architecture was useful to our research, design, and engineering: with it as a guide, we have created the Fawkes elastic MapReduce system~\cite{DBLP:conf/sigmetrics/GhitYIE14}.

However useful, our original reference architecture has important limitations. 
How to include in it portals, Software-as-a-Service, and other application-level approaches where the users of the ecosystem barely need to know of its existence to conduct their work?
How to include in it in-memory distributed file systems, and other software-based data management systems that span the memory,  network, and storage boundaries?
How to include in it the various DevOps tools? 

To address these questions, during the course of 2016 we have significantly revised
the reference architecture and extended it to cover the entire datacenter ecosystem. The new architecture\footnote{We gave the first public talks on our new reference architecture in November 2016, at ICT with Industry 2016 and, the same day, to a plenary session of the Lorentz Center Highlights, \url{http://www.lorentzcenter.nl/LCHighlights/abstracts.php?abstract=Iosup}.} includes the layers in the original reference architecture, plus a variety of other layers and a new, broader structure.
In this reference architecture, there are five core layers, 
(5) {\it Front-end} for the application-level functionality,
(4) {\it Back-end} for task, resource, and service management on behalf of the application,
(3) {\it Resources} for task, resource, and service management on behalf of the cloud operator,
(2) {\it Operations Service} for basic services that are typically associated with (distributed) operating systems, and
(1) {\it Infrastructure} for managing physical and virtual resources.
An orthogonal layer, (6) {\it DevOps}, covers functions essential to operating the datacenter but orthogonal to the service provided to customers, such as monitoring, logging, and benchmarking.
The sub-layering in Layers 4 and 5 helps classify the many emerging systems with finer granularity, and highlights the intense specialization that is currently emerging in this part of the ecosystem.
Since 2016, we have mapped to the new reference architecture a large number of well-known industry ecosystems (e.g., Google, Facebook, Uber, Netflix, the broad collection of Apache projects). Our experience suggests the reference architecture does encompass these industry ecosystems. 

We emphasize the difference between the two reference architectures through the example of a big data ecosystem based on MapReduce.
As Figure~\ref{fig:refarchidc} shows, the core ecosystem maps well to both our reference architectures. 
High-Level languages like Pig and Hive are based on the Map-Reduce programming model. Execution and Runtime management are left to Hadoop and HDFS, that distribute and execute Map-Reduce jobs. At a lower level, general-purpose resource allocation and scheduling in the datacenter is performed by Yarn or Mesos. Specific operations like the maintenance of configuration information for the upper layers can be performed by Zookeeper. 
This representation does not include the whole complexity of an industry datacenter stack, where there can be hundreds of additional components. 
Moreover, the old architecture, depicted in Figure~\ref{fig:refarchidc}~(top), does not capture in-memory file systems such as MemEFS~\cite{DBLP:journals/fgcs/UtaDWOSCK18} and Pocket~\cite{DBLP:conf/osdi/KlimovicWSTPK18}, high-performance network and storage engines such as Crail~\cite{DBLP:journals/debu/StuediTPSMIK17} and FlashNet~\cite{DBLP:journals/tos/TrivediIMSPKKG18}, DevOps tools such as Graphalytics~\cite{DBLP:journals/pvldb/IosupHNHPMCCSAT16} and Granula~\cite{DBLP:conf/grades/NgaiHHI17}, etc.

%% file: content/exp-serverless.tex
\begin{table}[!t]
  \caption{Co-evolving problem-solutions in our serverless work.}
  \label{tab:exp:serverless}
    \begin{adjustbox}{max width=\columnwidth}
  \begin{tabular}{lllll}
    \toprule
    Study & Feature & Depth & Breadth & W/ Team\\
    \midrule
    \multicolumn{5}{c}{{\it Longitudinal studies}}\\
    \midrule
    \cite{DBLP:journals/internet/EykTTVUI18} ('18) & Evolution & Deep & Narrow & SPEC RG Cloud \\
    \midrule
    \multicolumn{5}{c}{{\it Other studies}}\\
    \midrule
    \cite{conf/middleware/EykIST17} ('17) & General & Low & Wide & SPEC RG Cloud \\
    \cite{DBLP:conf/wosp/EykIAGE18} ('18) & Performance & Low & Wide & SPEC RG Cloud \\
    \midrule
    \multicolumn{5}{c}{{\it Instruments}}\\
    \midrule
    GitHub ('17-'19) & Fission WF. & Deep & Wide & Platform9 \\
    \cite{specrgcloud-serverless-ref-arch} ('19) & Ref. Arch & Low & Wide & SPEC RG Cloud \\
  \bottomrule
\end{tabular}
\end{adjustbox}
\end{table}

\subsection{Serverless: New Designs for FaaS}
\label{sec:exp:serverless}

We present in this section an overview of our design work in serverless computing, which emphasizes two aspects related to using the \AtLarge{} design framework in practice. 
First, approaching the rapidly evolving model of serverless computing as a co-evolving problem-solution enabled us 
to quickly gain insight into this domain.
Second, the \AtLarge{} research team has joined for this work with a variety of designers from academia and industry: a distributed team with background in performance engineering from the SPEC RG Cloud Group, another distributed team formed with serverless company Platform9, and, for control, by a team from Stanford and IBM Research Zurich working on serverless independently. 
The latter 
tests the ability of the \AtLarge{} design framework to help designers with different intellectual backgrounds and design approaches work together.
Table~\ref{tab:exp:serverless} summarizes our key contributions in this emerging field.

Serverless computing is part of a trend toward applications composed of many small, self-contained, and automatically managed components~\cite{DBLP:journals/internet/EykTTVUI18}.
{\it Serverless computing} is a set of (cloud) computing technologies that adhere to three principles\cite{conf/middleware/EykIST17}: 
(1) operational logic is abstracted away from the users; 
(2) users only pay for the resources they need, with fine granularity; 
(3) the computing model is event-driven and operations are scaled elastically.
Core to serverless computing, Function-as-a-Service (FaaS) allows developers to provide functions, for which the entire operational life-cycle is managed by the cloud provider. 

In early 2017, during an investigation into improving orchestration of container-based micro-services, we learned about the early industry efforts in serverless computing.
After an informal survey, we realized the benefits of this novel paradigm. Whereas micro-service architectures delegate most of the operations effort to the developer, the serverless model delegates most of the operational life-cycle to the cloud provider. This opens up many avenues for design to leverage this additional insight and control.
Following the \AtLarge{} design framework, we shifted our focus from improving micro-service architectures to the new problem of understanding the benefits and drawbacks of serverless computing.

As with any emerging domain led by industry efforts, serverless computing suffered (and, in some respects, still suffers) from a lack of a rigorous and scientific foundation: What is the definition of serverless? What are the characteristics of serverless technologies? What are the challenges and perspectives? How do these new technologies compare with each other and with traditional alternatives? How does it fit into or overlap with existing computing models?
To get answers to these questions, we established an international team within the SPEC RG Cloud Group, partnering both industry and academic organizations. 
In our initial publication, we addressed terminology, challenges, and perspectives~\cite{conf/middleware/EykIST17}---
key aspects for designers.

Following this initial high-level investigation, 
we narrowed the scope to the domain where we could leverage our existing expertise: that of performance, and of resource management and scheduling.
We then revisited and expanded the performance challenges introduced in the initial vision paper, identifying promising approaches towards addressing them~\cite{DBLP:conf/wosp/EykIAGE18}. 
To further separate hype from reality and leverage our decades of expertise in distributed systems, in another publication we reflected on the motivations, concepts, technologies, and practices that have led to the emergence of serverless computing \cite{DBLP:journals/internet/EykTTVUI18}. Our main finding was clear: though serverless technologies leverages and overlaps many historical efforts, its emergence could not have happened ten years ago. 

Continuing the exploration of serverless computing, we further narrowed our scope to the specific research challenges for which we could best leverage our (technical) expertise.
Within the SPEC RG Cloud Group, we focused on the core mission of the group: to perform quantitative system evaluation and analysis in distributed (eco)systems.
Towards a representative benchmark of serverless platforms, we spent a year surveying nearly 50 open-source and closed-source serverless(-like) platforms. As a culmination of these efforts, in early 2019 we proposed a FaaS reference architecture and ecosystem
that identifies the common processes and components in these seemingly widely varying systems. These processes and components are the focus of any good benchmark design for serverless computing.

Alongside this community effort, the AtLarge team started addressing the technical challenges associated with serverless computing, leveraging our expertise to introduce workflow-based serverless orchestration and serverless big data processing.
For example, we have created an informal public-private partnership, working with US-based company Platform9 to develop a production-ready serverless computing platform. Specifically, we have co-created the Fission Workflows\footnote{\url{https://github.com/fission/fission-workflows}} system, which acts as a workflow execution engine in the hierarchical Kubernetes-Fission ecosystem.

That the \AtLarge{} team is further exploring a (shared) storage architecture for serverless computing. As with any emerging field, the current serverless landscape gives us the opportunity to re-evaluate many of the basic design decisions (and trade-offs) present in current designs. 
For this, the \AtLarge{} team has been joined by a new high quality designer, who has already developed a recognized body of work in serverless computing with another team spanning Stanford and IBM Research Zurich~\cite{DBLP:conf/osdi/KlimovicWSTPK18,DBLP:conf/usenix/KlimovicWKSPT18}.
Their approach to design is compatible with the \AtLarge{} problem-solving process (Figure~\ref{fig:design:problemsolving}). First, the team identified the problem, formulated the new requirements for temporary storage for serverless, and analyzed the available trade-offs~\cite{DBLP:conf/usenix/KlimovicWKSPT18}. Then, they designed a complete system, with both high- and low-level components, and analyzed through detailed experiments the various design decisions and if they met the original objectives~\cite{DBLP:conf/osdi/KlimovicWSTPK18}.

Our work in serverless computing is just starting.
With numerous open challenges in this space, our efforts continue to identify evolving patterns, combine fundamental distributed systems notions in the emerging ecosystem of serverless technologies, and evaluate existing state-of-the-art systems to identify further pragmatic problems.

%% file: content/exp-graphalytics.tex
\begin{table}[!t]
  \caption{The emergence of the Graphalytics ecosystem. (Acronyms: NYP--not yet published, HPD--heterogeneous distributed and parallel system.)}
  \label{tab:exp:graphalytics}
  \begin{adjustbox}{max width=\columnwidth}
  \begin{tabular}{lllll}
    \toprule
    Study & Feature & Depth & Breadth & Released artifacts\\
    \midrule
    \cite{DBLP:conf/ipps/GuoBVIMW14} ('14) & PAD Law & Low & Wide & --- \\
    \cite{DBLP:conf/cluster/UtaVMLI18} ('18) & HPAD Law & Low & Wide & --- \\
    \midrule
    \cite{DBLP:conf/sigmod/CapotaHIPEB14} ('14) & PAD & Low & Wide & Graphalytics 0.1\\
    \cite{DBLP:journals/pvldb/IosupHNHPMCCSAT16} ('16) & PAD & Low & Wide & Graphalytics 1.0 \\
    \cite{DBLP:conf/grades/NgaiHHI17} ('17) & Sharing & Deep & Wide & Granula 0.1 \\
    NYP ('18) & PAD & Low & Wide & Global Competition \\
    \cite{tr:Grade10} ('18) & Modeling & Deep & Wide & Grade10 0.1 \\
    \midrule
    \cite{DBLP:conf/ccgrid/GuoVIE15} ('15) & GPUs & Deep & Narrow & --- \\
    \cite{DBLP:conf/ccgrid/GuoVEI16} ('16) & HPD & Deep & Narrow & --- \\
    \cite{DBLP:conf/ccgrid/AuUII18} ('18) & Elasticity & Deep & Wide & --- \\
  \bottomrule
\end{tabular}
\end{adjustbox}
\end{table}

\subsection{Design of the Graphalytics Ecosystem} \label{sec:exp:graphalytics}

We present in this section, which appears ad literam in our previous publication~\cite[$\S$6.1]{conf/hpdc/DesignProcess19}, an overview of the design of the Graphalytics ecosystem, which has emerged through multiple iterations of the \AtLarge{} design process.
The approach of co-evolving problem-solution has led to identifying new laws in the operation of graph-processing systems, to the development of an ecosystem of performance instruments and tools, and to meaningful and novel research directions. We discuss each in the following, and summarize the iterations in Table~\ref{tab:exp:graphalytics}.

Intrigued by a seminal analysis of open challenges in graph processing at scale~\cite{DBLP:journals/ppl/LumsdaineGHB07}, we have started planning to conduct our own experiments to understand the real-world problems. 
We designed experiments focusing on multi-algorithm, multi-dataset analysis of a diverse set of graph-processing platforms, exploring the 
dependency of performance on the interaction between software-platform, algorithm, and dataset (\emph{the PAD triangle}). 
It took us several years to get this curiosity-driven project done~\cite{DBLP:conf/ipps/GuoBVIMW14}. 

Finding that the PAD triangle existed ({\it a law}!) led us to a new problem, of providing the community with a benchmark that would not only support multiple ``P''s, as the leading benchmarks already did at the time~\cite{DBLP:books/sp/18/BonifatiFHI18}, but also multiple 
``A''s and ``D''s.
This bootstrapping led to the Graphalytics benchmark, the first tool in the emerging Graphalytics ecosystem of performance instruments and tools. 
The first solution was Graphalytics 0.1~\cite{DBLP:conf/sigmod/CapotaHIPEB14}, which is an engineered version of a subset of the initial PAD study---it takes much implementation effort to convert a scientific prototype into a production-ready software package.
With community involvement, and in particular the collaboration of the LDBC community, we have continued this line of design, to Graphalytics 1.0~\cite{DBLP:journals/pvldb/IosupHNHPMCCSAT16}. 
Graphalytics allowed us to benchmark, in production and in the lab, over ten graph-processing platforms. This has led us to new problems: How to enable a global competition around the benchmark? How to share performance data? 
How to enable not only low-depth analysis, which is typical of 
benchmarks, but also deep results? How to use the deep results to obtain model 
systems, without (much) 
effort? Table~\ref{tab:exp:graphalytics} indicates our incipient answers to these questions. Equally important, the Graphalytics artifacts have become a part of the official LDBC benchmarks, and serve a growing community of industry developers of graph-processing systems.

In parallel with tool-related problems, the Graphalytics ecosystem has spurred two new research directions. 
First, 
we have recently shown~\cite{DBLP:conf/cluster/UtaVMLI18} 
that for 
graph-processing platforms based on 
modern heterogeneous ``H''ardware the entire HPAD performance-space is relevant; 
the PAD law is applicable only in special situations. 
Second, understanding the performance impact of various emerging features in 
graph-processing. We have been some of the first to explore 
the performance of graph-processing platforms that are
(i) GPU-based~\cite{DBLP:conf/ccgrid/GuoVIE15}, 
(ii) based on parallel and distributed systems combined into a single working system~\cite{DBLP:conf/ccgrid/GuoVEI16}, 
(iii) elastic~\cite{DBLP:conf/ccgrid/AuUII18}.


%% file: content/exp-portfolio.tex
\begin{table}[!t]
  \caption{Characteristics of portfolio scheduling studies. Each study has a generally positive result, but also leads to a new research question.  (The acronyms, (i) general: PS--portfolio scheduling, (ii) for workloads (W): CE--Computer engineering, BC--business-critical workloads, BD--Big data, G--gaming, Ind--Industrial IoT analytics, Sci--scientific, Syn--synthetic,  (iii) for environment (Env): CL--own cluster, CD--public cloud, G--grid, GDC--Geo-distributed datacenters, MCD--multi-cluster datacenters.)}
  \label{tab:exp:portfolio}
  \vspace*{-0.35cm}
  \begin{adjustbox}{max width=\columnwidth}
  \begin{tabular}{lccll}
    \toprule
    Study & W & Env & Finding: PS is... & New Question\\
    \midrule
    \cite{DBLP:conf/jsspp/DengVRI13} ('13) & Syn & CL & useful & Works online?\\
    \cite{DBLP:conf/sc/DengSRI13} ('13) & Sci & G+CD & good online & Other W/Env?\\
    \cite{DBLP:conf/europar/ShenDIE13} ('13) & Sci+Gam & CL & useful & Other W/Env?\\
    \cite{DBLP:conf/jsspp/ShaiSF13} ('13) & CE & GDC & useful & Other W/Env?\\
    \cite{DBLP:journals/computer/BeekDHHI15} ('15) & BC & MCD & useful & Other W/Env?\\
    \cite{DBLP:conf/icac/MaISI17} ('17) & Ind & CD & useful & Other W/Env?\\
    \cite{DBLP:conf/bigdataconf/VoineaUI18} ('18) & BD & Cl & useful, but... & BD limits?\\
  \bottomrule
\end{tabular}
\end{adjustbox}
\end{table}

\subsection{Design of Portfolio Schedulers} \label{sec:exp:portfolio}

We present in this section, which appears ad literam in our previous publication~\cite[$\S$6.2]{conf/hpdc/DesignProcess19}, an experiment in using the \AtLarge{} design process to design datacenter schedulers.
We started with a series of comprehensive experiments about the performance of online job schedulers in grid datacenters---for BoT-~\cite{DBLP:conf/hpdc/IosupSAE08} and workflow-based~\cite{DBLP:conf/hpdc/SonmezYAIE10} workloads, for the predictive component of proactive schedulers~\cite{DBLP:conf/hpdc/SonmezYIE09}. The main conclusion across all the studies was that no individual technique or policy was consistently better than all others. 
A new need emerged, to select (change) the policy online, based on the current system state. 
This led us methods to select one policy among many,  
and ultimately to introduce portfolio scheduling in datacenters. Table~\ref{tab:exp:portfolio} captures this research development, which starts from a re-framing of the scheduling problem, triggered by a phenomenon found empirically.

We started with an exploration of the capabilities of portfolio scheduling across synthetic workloads with various computational properties~\cite{DBLP:conf/jsspp/DengVRI13}. While conducting this investigation, we found a new problem: the time it took a portfolio scheduler to simulate all the alternatives could grow rapidly, proportionally with the number of policies. 
Compounding this problem, BoT- and workflow-based workloads are comprised of many more jobs in the same time-span than traditional parallel workloads, a phenomenon not predicted by theory and that we had found around the same time~\cite{DBLP:journals/internet/IosupE11}; this means that simulators would have more to compute than predicted for previous approaches to (dynamic) online scheduling~\cite{DBLP:journals/software/FeitelsonN99}.
Thus, the portfolio scheduler could no longer be used to run online. This is an example of how solving an existing problem (of making a scheduler address system dynamics) can lead to a new problem (of making a scheduler fast enough to work online).

We thus turned our attention to (1) problems of real-world online scheduling, for (2) real workloads, here, scientific computing. 
This has led us to design a new portfolio-scheduling approach~\cite{DBLP:conf/sc/DengSRI13}, which could select a limited set active set of policies. The key trade-off in this design is keeping the active set large enough to make good decisions, yet small enough to 
estimate online.

But is portfolio scheduling generally capable? 
In successive iterations, we have tested portfolio scheduling on a variety of workloads and environments, which Table~\ref{tab:exp:portfolio} summarizes.
Portfolio scheduling seems indeed general, but requires non-trivial adaptation to workload and environment.
%
Independently, others have found portfolio scheduling useful for compute farms at Intel~\cite{DBLP:conf/jsspp/ShaiSF13}; this supports our claim that the \AtLarge{} design process can lead to meaningful designs, but also emphasizes that (i) it gives a relatively small research team the ability to compete intellectually with larger R\&D teams, and (ii) it led the research team to deeper and broader designs, consequence of the focus on the co-evolving problem-solution. 

Are there open problems?
Our latest study~\cite{DBLP:conf/bigdataconf/VoineaUI18}, on cluster-based big data workloads, indicates portfolio scheduling can make sub-optimal selections when the performance of the policy is difficult to predict. 
How to alleviate this problem remains open.

%% file: content/exp-autoscaling.tex
\subsection{Design of Autoscaling Experiments} \label{sec:exp:autoscaling}

We present in this section, which appears ad literam in our previous publication~\cite[$\S$6.3]{conf/hpdc/DesignProcess19}, an overview of using the \AtLarge{} experiment design for experiments on autoscaling in cloud environments. 
Autoscaling systems try to provision exactly as many resources as the workload demands, by provisioning on behalf of the user more or fewer resources. 
An autoscaler is an algorithm used by an autoscaling system to automate elasticity efficiently, subject to one or several common elasticity metrics~\cite{DBLP:journals/tompecs/HerbstBKOEKEKBA18}.
Overall, the process allowed us to conduct successful and deep experiments. The various aspects addressed in this section indicate both how complex the experiment design needs to be, to address various aspects of distributed systems and ecosystems, and that the process fosters successful experiment designs.

Key to our work in autoscaling was to understand how autoscalers perform in practice, for the emerging class of {\it workflow-based} cloud workloads. When we started this work, there existed already several autoscaling approaches, but none had been evaluated comprehensively. Motivated also by the emergence of a new set of elasticity metrics, we have designed and performed several experiments~\cite{DBLP:conf/wosp/IlyushkinAHPGEI17,journal/tompecs/IlyushkinAHPEI18,DBLP:conf/ccgrid/VersluisNI18}.

For the first set of experiments~\cite{DBLP:conf/wosp/IlyushkinAHPGEI17}, we have designed a new morphological structure for autoscaling workflows, based on general and workflow-specific autoscalers. We have further selected in vitro emulation as the evaluation technique, ten elasticity metrics, various environment and workload configurations. For our experiments, we have developed real-world software, implemented real-world policies, and custom experimental tools to run this in the DAS multi-cluster system~\cite{DBLP:journals/computer/BalELNRSSW16} set to emulate a cloud. Last, we have designed and conducted $N=5$ experiments, and further designed two ranking methods to aggregate the results into head-to-head comparisons---{\it which policy is the best?} 

Although the in vitro experiments were useful, they could not address many questions related to diverse workloads and environments, because the former would have been too expensive to experiment with, and for the latter we did not have access to different environments (and of the right scale). We have therefore designed and conducted in silico, simulation-based experiments~\cite{DBLP:conf/ccgrid/VersluisNI18}. 
We found interesting discrepancies between the real-world software of the initial in vitro experiments and the software of the simulator, which we have developed independently; these discrepancies have allowed us to correct in time the real-world results, and emphasize the need for {\it independent corroboration} in the community~\cite{experimental:Feitelson06}.

We have then extended this work with more comprehensive analysis of the results~\cite{journal/tompecs/IlyushkinAHPEI18}. The new analysis exemplifies the depth of stage (9) in the \AtLarge{} approach: we added an analysis of traditional performance metrics next to the analysis of elasticity metrics, an analysis of cost metrics based on several real-world cost models, an analysis of introducing two types of deadline-based SLAs, and an analysis of the presence of performance variability in the behavior of autoscalers. We have also introduced a method to grade autoscalers, by combining their scores judiciously. 


%% file: content/01-relatedwork.tex
\section{Related Work}
\label{sec:related}

In this section, we compare our and related work.

{\bf Overall novelty:} 
The \AtLarge{} design framework combines elements of 2010s design thinking with the specifics of \ourdomainshort{} design. 
The former makes it unique among published design frameworks in distributed systems. For example, hardware design is a well-established field of design, but as noted by Brooks it has not adopted the new ways of design thinking~\cite[Part I]{design:book/Brooks10}.
The latter makes it unique among design frameworks. For example, 
works of similar scope address the design of mechanical systems~\cite{design:book/BeitzPFG07,book:design:Ullman17}, but their physical properties makes them radically different from distributed systems and ecosystems.

{\bf Contrast to design in computer systems:}
We distinguish here two design cultures.
{\it Hardware design} has focused for over five decades on (instruction set) architecture as function, implementation of the system to solve in particular for cost-performance among NFRs, and realization to engineer the working system~\cite{book:design:Buchholz62,design:principle:parsimony73,book:design:BlaauwB97,book:compsys:HennessyP17}. 
Standardization and increased capabilities of simulation software has made design space exploration largely computer-driven, focusing on the optimization of fixed design-spaces. 
Post-Moore's Law, 
we have seen a wave of innovative hardware designs, including heterogeneous CPUs (e.g., second-generation KNL), GPUs, FPGAs, and various ASICs, but so far the emergence of a new design process has not been reported.
{\it Software system design} has been much less developed, caught perhaps between software engineering and hardware design. Thus, this area of design has focused mostly on reporting best-practices and rules-of-thumb for addressing NFRs, such as 
scalability~\cite{book:design:Abbott11}, various other NFRs in cloud-based ecosystems~\cite{book:design:Abbott15,book:design:Burns18}; and pragmatic operational issues~\cite{book:design:Nygard07,book:design:Keeling17,book:google:SRE16,book:google:SRW18}.

The \AtLarge{} design process is not closely aligned with either of these approaches; 
following the field-wide critique of Brooks~\cite[Part I]{design:book/Brooks10}, it focuses on co-evolving problem-solutions, problem-solving {\it and} problem-finding, etc.
For example, 
the traditional principles of system design~\cite{design:principle:parsimony73,book:design:BlaauwB97,book:compsys:HennessyP17} are not the same as the principles we propose for \ourdomainshort{}, and the \AtLarge{} approaches to problem-finding and problem-solving are distinctively more systematic than the published best-practices of software system design.

{\bf Contrast to design in software engineering:}
Software engineering has developed and keeps evolving sophisticated design methods~\cite{DBLP:journals/csur/RamsinP08}. 
Elements of software design provide 
various analytical views~\cite{Rozanski2005,book:design:Bass12},
software-oriented design patterns as problem-solution recipes~\cite{book:design:GOF94,book:design:Alexandrescu01},
documentation~\cite{book:design:Clements10} and maintenance of code, DevOps from a Dev's perspective~\cite{concepts:book/Bass15},
etc.
In contrast to these approaches, the \AtLarge{} design process focuses on {\it systems}.

{\bf Contrast to design in general:}
To design our process, we have surveyed design processes and elements from mechanical engineering~\cite{design:book/BeitzPFG07,book:design:Ullman17}, operations research and management~\cite{book/Simon96}, architecture~\cite{book:design:Alexander77,book:design:Rowe91,book:design:Rybczynski13}, material and fashion design~\cite{book:design:Aspelund14}, 
graphic design~\cite{book:design:Meggs16}, 
industrial and facility design~\cite{book:tech:Freeman18}, etc.
In contrast to these approaches, the \AtLarge{} design process
provides different solutions due to the virtual, composite, and idiosyncratic nature of the distributed systems and ecosystems.

%% file: content/01-conclusion.tex
\section{Conclusion}
\label{sec:conclusion}

Responding to the needs of an increasingly digital and knowledge-based society, in this work we explicitly posit that design is a key area of research for distributed systems and ecosystems~(\ourdomainshort{}), and propose a vision to establish the theory and practice of \ourdomainshort{}~design.

We propose the first attempt to understand the problem of \ourdomainshort{} design. We give qualitative and quantitative evidence of the extent of the problem, and propose requirements derived from general design processes and from the specific needs of \ourdomainshort{}.

We design the \AtLarge{} design framework around the central premise that design is fundamentally different from science and engineering, requiring its own way of thinking and processes. Responding to requirements, the framework combines emerging theories about design thinking with several \ourdomainshort{}-focused design processes, e.g., for co-evolving problem-designs, for problem-finding and -solving, and for disseminating the results. 

We show how, in our experience, the framework can lead to pragmatic and innovative designs in fields such as 
P2P systems, 
datacenter ecosystems, 
ecosystems for the MMOG application-domain, 
serverless computing and FaaS cloud computing, 
DevOps ecosystems for performance analysis,
system-level design of a portfolio scheduler for datacenters,
and experiment design for analyzing autoscaling,
etc.

Our vision also includes a set of core principles and challenges of \ourdomainshort{} design, in the four broad categories related to the central premise, systems, peopleware, and method.
We have started to address the research agenda formulated in this article. We hope this vision will stimulate a larger community to join us in improving design.

%% file: content/00-ack.tex
\section*{Acknowledgments}\vspace*{-0.25cm}
{\small
	Work supported 
	by the projects Vidi MagnaData and Commit.
	We thank all our collaborators, in particular, in the SPEC RG Cloud Group, at TUD, at VU and UV Amsterdam, at Platfom9, at Oracle, at Intel Labs, etc.
}